\documentclass[10pt]{article}
\usepackage[T1]{fontenc}
\usepackage{times}
\usepackage{authblk}
\usepackage{amssymb,amsmath,wasysym}
\usepackage{booktabs}
\usepackage[letterpaper,margin=1in]{geometry}
\usepackage{graphicx}
\usepackage{natbib}
\usepackage{enumitem}
\usepackage{capt-of}
\usepackage[compact]{titlesec}
\usepackage{setspace}

\title{Highly Siderophile Elements in the Earth's Mantle as a Clock for the Moon-forming Impact}
\date{}
\author[1,2]{Seth A. Jacobson\thanks{seth.jacobson@oca.eu}}
\author[1]{Alessandro Morbidelli}
\author[3,4]{Sean N. Raymond}
\author[5]{David P. O'Brien}
\author[6]{Kevin J. Walsh}
\author[2]{David C. Rubie}
\affil[1]{Observatoire de la C{\^o}te d'Azur, Laboratoire Lagrange, 06304 Nice, France}
\affil[2]{Universt{\"a}t Bayreuth, Bayerisches Geoinstitut, 95440 Bayreuth, Germany}
\affil[3]{Universite Bordeaux, Laboratoire d'Astrophysique de Bordeaux, UMR 5804, F-33270 Floirac, France}
\affil[4]{CNRS, Laboratoire d'Astrophysique de Bordeaux, UMR 5804, F-33270 Floirac, France}
\affil[5]{Planetary Science Institute, 1700 E. Ft. Lowell, Suite 106, Tucson, AZ 85719, USA}
\affil[6]{Southwest Research Institute, Planetary Science Directorate, 1050 Walnut St., Suite 300, Boulder, CO 80302, USA}

\begin{document}
\maketitle
{\bf According to the generally accepted scenario, the last giant impact on the Earth formed the Moon and initiated the final phase of core formation by melting the Earth's mantle.
A key goal of geochemistry is to date this event, but different ages have been proposed.
Some argue for an early Moon-forming event, approximately 30~million years (Myr) after the condensation of the first solids in the Solar System~\citep{Yin:2002ex,Jacobsen:2005bj,Taylor:2009jr}, whereas others claim a date later than 50~Myr (and possibly as late as around 100~My) after condensation~\citep{Touboul:2007is,Allegre:2008iq,Halliday:2008bo}.
Here we show that a Moon-forming event at 40~Myr after condensation, or earlier, is ruled out at a 99.9 per cent confidence level.
We use a large number of N-body simulations to demonstrate a relationship between the time of the last giant impact on an Earth-like planet and the amount of mass subsequently added during the era known as Late Accretion.
As the last giant impact is delayed, the late-accreted mass decreases in a predictable fashion.
This relationship exists within both the classical scenario~\citep{Obrien:2006jx,Raymond:2009is} and the Grand Tack scenario~\citep{Walsh:2011co,OBrien:2014bk} of terrestrial planet formation, and it holds across a wide range of disc conditions.
The concentration of highly siderophile elements (HSEs) in Earth's mantle constrains the mass of chondritic material added to Earth during Late Accretion~\citep{Chyba:1991cq,Bottke:2010hr}. 
Using HSE abundance measurements~\citep{Becker:2006bi,Walker:2009be}, we determine a Moon-formation age of 95~$\pm$~32~Myr after the condensation.
The possibility exists that some late projectiles were differentiated and left an incomplete HSE record in Earth's mantle.
Even in this case, various isotopic constraints strongly suggest that the late-accreted mass did not exceed 1 per cent of Earth's mass, and so the HSE clock still robustly limits the timing of the Moon-forming event to significantly later than 40~My after condensation.}

The Moon-forming impact must be the last giant impact experienced by Earth, because both Earth and the Moon share an almost identical isotopic composition.
This giant impact melted Earth's mantle and formed the final global magma ocean, causing core-mantle differentiation, and it possibly removed a significant portion of Earth's atmosphere.
These events can be dated using radiometric chronometers. 
Unfortunately, the age of the Moon differs substantially from one chronometer to the next owing to assumptions in the computation of the so-called model ages.
For instance, \citet{Touboul:2007is} measured very similar $^{182}$W/$^{184}$W ratios for both Earth and the Moon and, given the differences in Hf/W ratios estimated at the time for the two bodies, concluded that the Moon-forming event must have been 62$_{-10}^{+90}$ Myr after condensation.
(Through-out, we use `after condensation' to mean `since the birth of the Solar System'; see the Supplementary Materials for details.)
In this way, the radioactive $^{182}$Hf would have almost fully decayed into $^{182}$W beforehand, thus easily accounting for the almost non-existent difference in $^{182}$W/$^{184}$W ratios between the Moon and Earth.
However,~\citet{Konig:2011we} subsequently determined that the Hf/W ratios of the Moon and Earth are instead identical, voiding this reasoning and leaving the problem of dating the Moon-forming event wide open.

We approach this problem from a new direction, using a large number of $N$-body simulations of the accretion of the terrestrial planets from a disk of planetesimals and planetary embryos.
The simulations extend across the range of well-studied and successful scenarios (for example, typically creating the right number of planets on the correct orbits) and fall into two categories.
First, classical simulations assume that the disk of terrestrial building blocks extended from an inner edge at 0.3--0.7 AU (where 1 AU is the Earth-Sun distance) from the Sun out to 4--4.5 AU, just interior to Jupiter, and that the giant planets stayed on orbits near their current ones.
Our sample of 48 classical simulations produces 87 Earth-like planets~\citep{Obrien:2006jx,Raymond:2009is}, which are broadly defined as final bodies with masses within a factor of two of Earth's mass and with orbits between the current orbits of Mercury and Mars.
In these simulations, the last giant impacts occur between 10~Myr and 150~My after the removal of the solar nebular gas, which happened about 3~Myr after condensation.
The classical scenario has known shortfalls.
The Mars analogs are too massive unless the giant planets are assumed to be initially on very eccentric orbits~\citep{Wetherill:1991wc,Raymond:2009is}.
These large eccentricities, however, cannot be explained in the context of the formation and evolution of giant planets in a gas disk.
Furthermore, terrestrial planets accreted in the presence of very eccentric giant planets are very low in water content~\citep{Chambers:2002hj,Obrien:2006jx}.

The second category consists of simulations in which the terrestrial planets form from a disk with an abrupt outer edge at about 1 AU; the inner edge remains the same as in the classical simulations.
Simulated solar systems in this second category successfully form Mars-like planets~\citep{Hansen:2009ke,Walsh:2011co,OBrien:2014bk}.
The truncation of the outer edge can be explained by the early gas-driven inward-then-outward migration of Jupiter and Saturn known as the Grand Tack, which then produces a region of greatly depleted surface density between the current orbits of Mars and Jupiter~\citep{Walsh:2011co}.
These simulations best reproduce the orbital and mass distributions of the terrestrial planets, and they also explain the compositional structure of the asteroid belt~\citep{Walsh:2011co}.
Moreover, Earth-like planets accrete volumes of water that are consistent with estimates of the Earth's water content~\citep{OBrien:2014bk}.
Previously reported Grand Tack simulations feature last giant impacts occurring typically within about 50~Myr after the time of removal of the solar nebular gas~\citep{Hansen:2009ke,Walsh:2011co,OBrien:2014bk}.
We complement those simulations with new ones (resulting in a total of 211 Grand Tack simulations producing 354 Earth-like planets).

All Grand Tack simulations produce planetary systems that match the Solar System as well as or better than those obtained from classical simulations despite varying many initial disc conditions (see Extended Data Figure 1), among them the initial total mass ratio between the embryo and planetesimal populations (from 1:1 to 8:1, see Supplementary Information).
If the initial ratio of embryo mass to planetesimal mass is increased, the time of the last giant impact also increases (even to about 150~Myr after condensation, see Extended Data Figure 2) owing to the reduction of the well-known effect of dynamical friction---the damping of the eccentricities and inclinations of the larger bodies due to gravitational interactions with a swarm of smaller bodies~\citep{Obrien:2006jx}.
Higher eccentricities and inclinations of the embryos diminish mutual gravitational focusing, increasing the accretion timescale, and consequently leading to later embryo-embryo collisions (that is, giant impacts). 

Dynamical models alone do not indicate whether the Moon-forming impact occurred early (about 30~Myr after condensation) or late (about 50--100~Myr after condensation), because the result depends on the initial disc conditions.
However, we find a clear statistical correlation between the time of the Moon-forming impact and the mass subsequently accreted, known as the late-accreted mass.
This era of Late Accretion includes no giant impacts by definition and so all of the late-accreted mass comes from the planetesimal population.
As shown in Figure 1, this correlation exists across all simulations of both types: classical and Grand Tack.
We interpret this correlation by considering that the planetesimal population decays over a characteristic time, so that if the last giant impact occurs earlier, then the remaining planetesimal population is larger.
A larger remaining planetesimal population delivers a larger late accreted mass.
Strengthening the correlation, a larger initial planetesimal population leads to a shorter timescale for giant impacts owing to enhanced dynamical friction.
For any given last-giant-impact time, Earth-like planets in the classical simulations acquire larger late-accreted masses than those in the Grand Tack simulations (see Figure 1), because the planetesimal population is more dispersed in the classical scenario and therefore decays more slowly. 

The correlation displayed in Figure 1 can be used as a clock that is independent of radiometric dating systems.
The late-accreted mass is input into this clock and the time of the last giant impact is read out.
A traditional estimate for the late-accreted mass can be obtained from the highly siderophile element (HSE) abundances in the Earth's mantle relative to the HSE abundances in chondritic meteorites~\citep{Chyba:1991cq,Bottke:2010hr}.
HSEs partition strongly into iron, and so are transported from the mantle to the core during core formation.
In this process, the element ratios are strongly fractionated relative to chondritic proportions~\citep{Mann:2012fp}.
The HSEs in Earth's mantle are significantly depleted relative to chondritic bodies---a clear consequence of core formation---and yet the remaining HSEs are in chondritic or near-chondritic proportions relative to each other~\citep{Becker:2006bi,Walker:2009be}.
This is commonly interpreted as evidence that all or a large portion of the HSEs currently in the mantle were delivered by chondritic bodies after the closure of Earth's core, an accretion phase known as the Late Veneer~\citep{Chou:1978uu}.
To account for the observed mantle budget of HSEs, we estimate that a chondritic mass of 4.8~$\pm$~1.6 $\times$~10$^{-3}$~M$_\oplus$ is necessary, where M$_\oplus$ represents an Earth mass.
This mass does include contributions from the era known as the Late Heavy Bombardment.
Current mass estimates for this very late (approximately 500~Myr after condensation) accretion are 10$^{-4}$~M$_\oplus$~\citep{Morbidelli:2012ko}, which we added to the late-accreted masses of our synthetic Earth-like planets, but it only accounts for about 2\% of the chondritic mass and therefore does not play a significant role in the analysis of the correlation.

The chondritic mass can only be identical to the late-accreted mass or to the Late Veneer mass if the Moon-forming event stripped all of the HSEs from Earth's mantle or was the last episode of growth for Earth's core, respectively (as is traditionally assumed).
However, these conditions are not necessarily true.
Consider that some projectiles colliding with Earth after the Moon-forming event might have been differentiated, so that their HSEs were contained in their cores.
If part of these cores had merged with Earth's core~\citep{Albarede:2013hd}, then the late-accreted mass would clearly be larger than the chondritic mass, because there would be no HSE record of this fraction of the projectile cores in Earth's mantle.
Additionally, in this case, given that iron (and therefore HSEs) would have been added to Earth's mantle and its core, the chondritic mass would be larger than the Late Veneer mass, which is geochemically defined as the mass accreted to Earth after the core has stopped growing.

In fact, as explained in detail in the Supplementary Information and Extended Data Figures 3 and 4, it is unlikely that more than 50\% of a projectile's core directly reaches Earth's core, otherwise geochemical models cannot reproduce the tungsten isotope composition of the Earth's mantle~\citep{Rudge:2010fv}.
Moreover, a large late-accreted mass, delivered in only a few objects so as to explain the relative HSE abundances of Earth and the Moon~\citep{Bottke:2010hr}, would have left a detectable isotopic signature on the Earth relative to the Moon, which is not observed~\citep{Wiechert:2001il,Zhang:2012dn}.
Thus, even when considering these more complex possibilities, geochemical evidence constrains the late-accreted mass probably not to exceed 0.01~M$_\oplus$ (see Supplementary Information).

For these reasons, we first make the usual assumption that the late accreted mass and the HSE-derived chondritic mass are identical.
In this case, not a single simulated Earth-like planet with a last giant impact earlier than 48~Myr since condensation has a late-accreted mass in agreement with the value estimated from HSEs (see Figure 1).
Of those forming in less than 48~Myr, only one planet is near the upper 1-$\sigma$ bounds of the chondritic mass.
Only after 67~Myr since condensation are there Earth-like planets with late-accreted masses consistently within the 1-$\sigma$ uncertainty bounds for the chondritic mass.
After 126~Myr since condensation, the late-accreted masses of Earth-like planets are often significantly below the lower limit set by the HSE measurements.

We calculate the log-normal mean and standard deviation of the late accreted masses of all Earth-like planets with last giant impacts within a range around a chosen time (see Figure 1).
We interpret these distributions as a model of the likelihood of a specific late accreted mass given a last giant impact time.
Given this likelihood model, we compute the fraction of Earth-like planets with late accreted masses not exceeding the HSE-derived chondritic mass, also taking into account the uncertainties on the latter.
Using only the Grand Tack simulations, which provide the best match to the terrestrial planet~\citep{Hansen:2009ke,Walsh:2011co,OBrien:2014bk}, the probability that an Earth-like planet with a last giant impact at or before 40~Myr since condensation has an HSE budget consistent with observations is 0.1\% or less (see Figure 2).
Furthermore, if we relax the assumption that the late-accreted mass is equal to the chondritic mass, but consider that the former can be up to 0.01~M$_\oplus$, then a Moon-forming event at or before 40~Myr is still ruled out at a 97.5\% confidence level or higher.

Before accepting the proposed HSE clock, a couple caveats need to be discussed.
First, our simulations always assume perfect accretion---that is, all mass from both colliding bodies ends up in the remaining planet.
Accounting for imperfect accretion---in which a fraction of the total mass is ejected away from the planet, the true late-accreted mass is probably smaller than that inferred from our simulations. 
Using the characteristics of each impact (for example, impact velocity and angle) recorded in our simulations and an algorithm developed from a large suite of numerical experiment~\citep{Leinhardt:2012kd}, we estimate the fraction of the projectile mass retained in each impact. 
The effect of imperfect accretion somewhat increases the likelihood of early giant impacts (see Figure 2). 
However, $N$-body simulations that have incorporated imperfect accretion showed that most of the mass not retained in the giant impacts is subsequently re-accreted~\citep{Kokubo:2010im,Chambers:2013cp}. 
Thus, our calculation incorporating imperfect accretion should be interpreted as an upper limit to the likelihood for a given last giant impact time. 
Reinforcing this analysis, if the HSEs are mostly delivered from the cores of differentiated projectiles as in the scenario of \citet{Bottke:2010hr}, imperfect accretion has a minimal effect on the HSE budget because the material that is lost into space comes predominantly from the projectile's mantle~\citep{Asphaug:2006gp}. 

Second, mutually catastrophic collisions between planetesimals may break them into ever smaller pieces until they are small enough to be removed by solar radiation before they can be accreted onto planets.
If this were a significant process, the late-accreted mass would be smaller than estimated from our simulations, which do not include this process.
However, the size distribution of craters on the lunar highlands suggests that---as in the current asteroid belt---most of the planetesimals' mass was in objects larger than 100 km~\citep{Strom:2005ur}.
The collisional comminution for bodies this large is negligible.
Moreover, most of the late-accreted mass was probably delivered by Ceres-sized (about 1,000 km across) or even larger bodies to explain the relative difference in HSE abundances between Earth and the Moon~\citep{Bottke:2010hr}.

Considering all of the above, we argue that the clock derived from the correlation between late-accreted mass and last-giant-impact age is robust and may be the most reliable way to estimate the age of the Moon-forming event.
Given the current constraints on the Earth's late-accreted mass, it establishes that this event occurred significantly later than 40~Myr and likely at 95~$\pm$~32~Myr after the formation of the first Solar System solids, in agreement with some~\citep{Allegre:2008iq,Halliday:2008bo} (but not all) previous estimates based on radiometric chronometers.
Future analyses will establish firmer constraints on the late-accreted mass, for instance, by determining the difference in W-isotope composition between Earth and the Moon more precisely. 
From this, we can obtain better limits on the timing of the Moon-forming event through our clock.

Our analysis of numerical simulations can also be used to support or invalidate different accretion scenarios.
For instance, \citet{Albarede:2013hd} consider the possibility of the impact with Earth of a few projectiles that have a total mass of 0.04~M$_\oplus$ as late as 130~Myr since condensation, but Figure 1 shows that it is extremely unlikely for so much mass to be delivered so late. 

A late Moon formation has at least two profound implications.
First, it constrains the dynamical conditions of the disk from which the planets accreted and the physical properties of the disk material.
For instance, a late last giant impact implies that most of the mass was in the embryo population rather than the planetesimal population. 
Second, reconciling a late Moon-forming event with radiometric chronometers that suggest the opposite result may require challenging fundamental assumptions, such as envisioning a Moon-forming event that did not reset all clocks simultaneously and left significant parts of the mantle non-equilibrated with the core.
This may argue in favor of some of the new scenarios proposed for the Moon-forming collision~\citep{Cuk:2012hj,Reufer:2012dz}, which distribute impact energy heterogeneously and may leave a significant portion of the Earth's mantle relatively undisturbed.

\begin{figure}
\includegraphics[width=\textwidth]{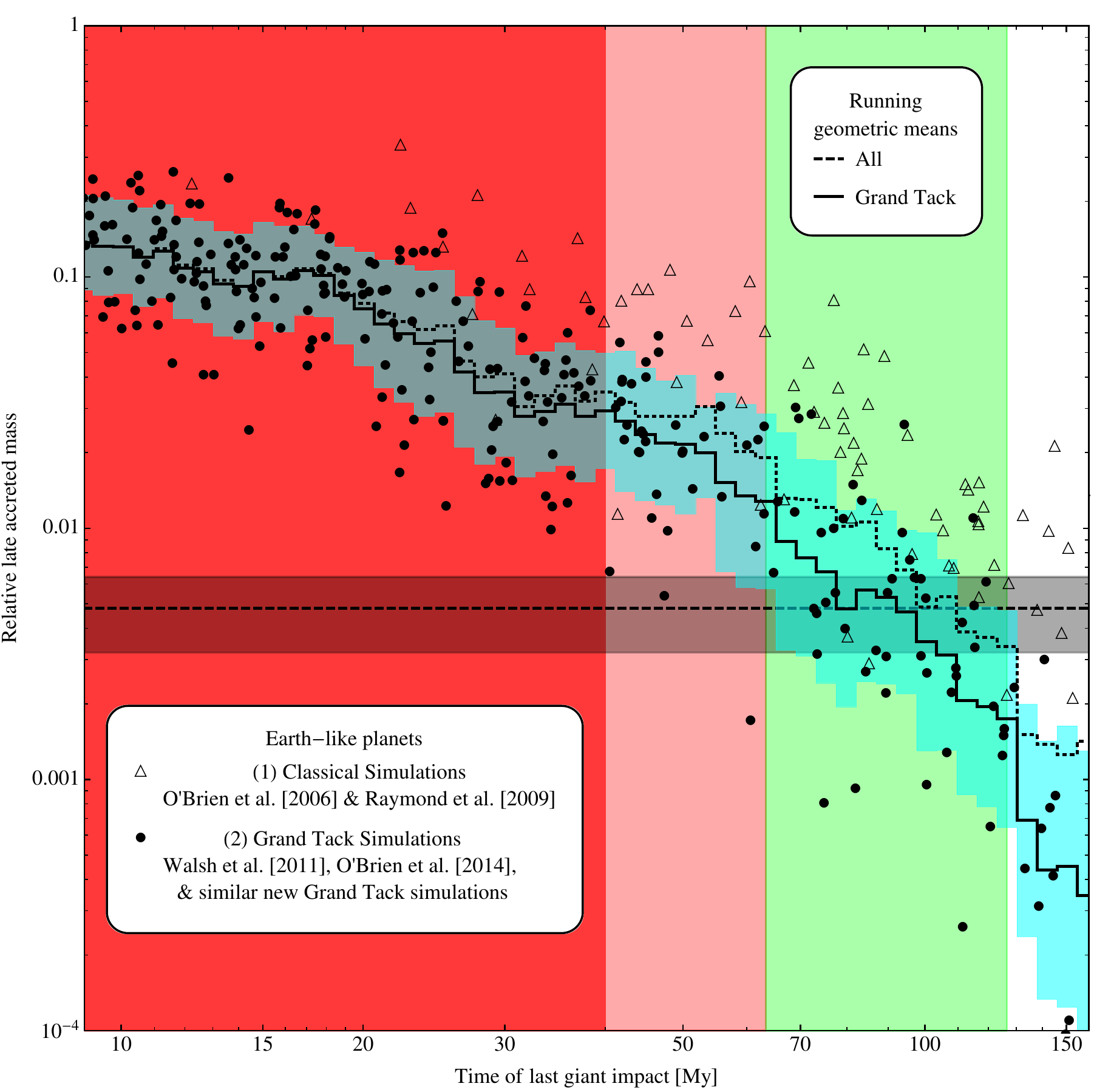}
\caption{The late-accreted mass relative to each synthetic Earth-like planet's final mass as a function of the time of the last giant impact. Triangles represent Earth-like planets from the first category: classical simulations with Jupiter and Saturn near their contemporary orbits~\citep{Obrien:2006jx,Raymond:2009is}. Circles represent Earth-like planets from the second category: Grand Tack simulations with a truncated protoplanetary disk~\citep{Walsh:2011co,OBrien:2014bk}. The black line resembling a staircase is the moving geometric mean of the late-accreted masses in the Grand Tack simulations evaluated at logarithmic time intervals with a spacing parameter of 0.025 and a width parameter twice that. The blue region encloses the 1-$\sigma$ standard deviation of the late-accreted mass, computed assuming that the latter is distributed log-normally about the geometric mean. Always predicting larger late accreted masses for each last giant impact time, the dotted staircase is the geometric mean obtained by also considering the classical simulations, although the latter do not fit Solar System constraints as well as the Grand Tack simulations do. The horizontal dashed line and enclosing darkened region are the best estimate and 1-$\sigma$ uncertainty of the late-accreted mass inferred from the HSE abundances in the mantle (chondritic mass): 4.8~$\pm$~1.6 $\times$~10$^{-3}$~M$_\oplus$. The best estimate for the intersection of the correlation and the chondritic mass is 95~$\pm$~32~Myr. The dark and light red regions highlight Moon-formation times that are ruled out with 99.9\% (40~Myr) and 85\% (63~Myr) confidence, respectively.}
\end{figure}

\begin{figure}
\includegraphics[width=\textwidth]{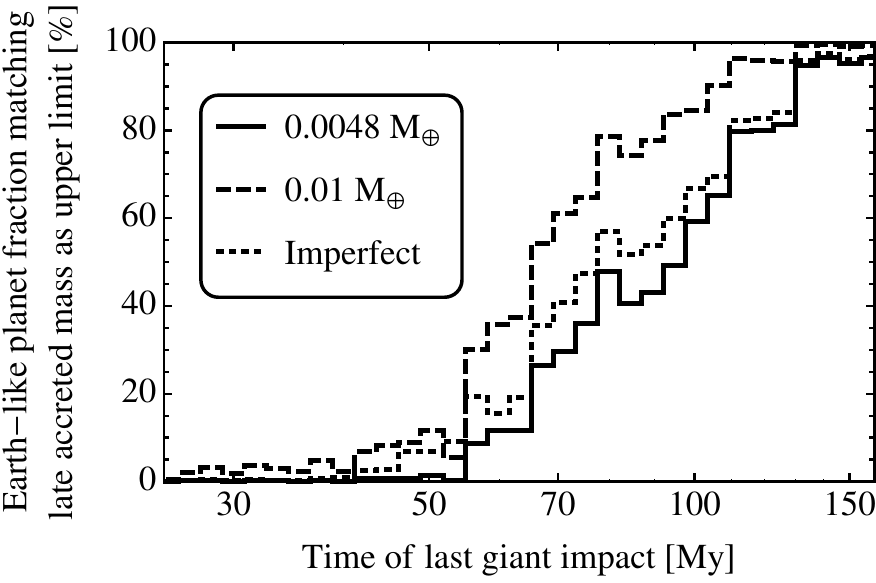}
\caption{The likelihood that a planet suffering a last giant impact within a specific range of times has a late-accreted mass less than or equal to the chondritic mass of 4.8~$\pm$~1.6 $\times$~10$^{-3}$~M$_\oplus$. The probability calculation uses the same bins as Figure 1 but only includes the Grand Tack simulations, because they best reproduce the terrestrial planets~\citep{Hansen:2009ke,Walsh:2011co,OBrien:2014bk}. The solid line shows this probability assuming perfect accretion and corresponds exactly to the late-accreted masses shown as circles in Figure 1. The lower 1-$\sigma$ limit for the Moon formation age is 63~Myr, which corresponds to a 15\% probability that an Earth-like planet with a last giant impact at that age is characterized by a late-accreted mass equal or smaller than the chondritic mass. The dashed line shows the same calculation but for a late-accreted mass less than or equal to 0.01~M$_\oplus$, which is an upper limit established from a number of elemental and isotopic constraints (see Supplemental Information). The dotted line shows the same calculation as for the solid line---that is, using a chondritic mass---but assuming imperfect accretion during collisions. This decreases the late-accreted masses by a variable amount depending on the impact characteristics of the late-accreted projectiles onto each planet~\citep{Leinhardt:2012kd}. However this calculation underestimates the late-accreted mass because a large fraction of the ejected material would be subsequently re-accreted~\citep{Kokubo:2010im,Chambers:2013cp} and projectile core material is less likely to be ejected post-impact~\citep{Asphaug:2006gp}. Consequently the dotted line overestimates the likelihood that a planet matches the chondritic mass constraint. A realistic estimate therefore lies between the solid and dotted curves (probably closer to the former).}
\end{figure}

\subsection*{Acknowledgements} S.A.J., A.M., D.P.O. and D.C.R were supported by the European Research Council (ERC) Advanced Grant "ACCRETE" (contract number 290568 to D.C.R.). D.P.O. was also supported by grant NNX09AE36G from NASA's Planetary Geology and Geophysics research program. S.N.R. thanks the CNRS's PNP program and NASA Astrobiology Institute's VPL lead team for their support.
\subsection*{Author Contributions} S.A.J., S.N.R., D.P.O. and K.J.W. ran numerical simulations and provided reduced data utilized in this study; S.A.J. analyzed results and discovered the relationship;  S.A.J., A.M., and D.C.R. discussed the results; S.A.J. wrote the paper with guidance from A.M.; All authors commented on the manuscript.

\part*{Supplementary Information}
\section{Numerical Simulations} The simulations within the main text come from four published sources~\citep{Obrien:2006jx,Raymond:2009is,Walsh:2011co,OBrien:2014bk} and a series of new Grand Tack simulations. They are all N-body simulations, which utilize either the Symba~\citep{Duncan:1998gn} or the Mercury~\citep{Chambers:1999jd} numerical integrator to calculate the gravitational forces between embryos and between embryos and planetesimals (for more details regarding simulations of terrestrial planet formation see a review by \citet{Morbidelli:2012iz}, and for the Grand Tack specifically, see the Supplementary Information of \citet{Walsh:2011co}). The simulations used in this work are broken into suites of runs with similar initial conditions. A suite of simulations consists typically of 10, but anywhere between 8 and 20, independent runs with similar initial protoplanetary discs. In the main text, we identified two categories of simulations that all of the suites fall into: (1) classical and (2) Grand Tack.

The classical model of terrestrial planet formation assumes that Jupiter and Saturn formed on or near their current orbits. They did not undergo any large-scale migration during the few million year gaseous disc phase and were fully-formed during the last phases of terrestrial accretion. The N-body simulations begin after the gas has been dispersed (i.e. $\sim$~3~My after the first solids in the Solar System~\citep{Haisch:2001bx}). The simulations used in the main text are taken from \citet{Obrien:2006jx} and \citet{Raymond:2009is}. The initial conditions of these simulations include solid bodies from an inner edge at 0.3--0.7 AU out to 4--4.5 AU.  They include a few tens to $\sim$~100 planetary embryos (Moon- to Mars-mass bodies) and $\sim$~1000--2000 planetesimals (Ceres-mass bodies) with an equal mass in each population.  The orbits of the giant planets are varied in different simulations, but Jupiter is always close to its current location.  The most important parameter that was changed in these simulations was the giant planets' eccentricities, which were tested between 0--0.1. Extended Data Figure 1 shows as open triangles the outcome of these simulations including the mass-orbit distribution, the timing of the last giant impact on Earth, the late accreted mass, and two orbital structure statistics.

We note that simulations using the classical model have consistently failed to reproduce the terrestrial planets' mass-orbit distribution~\citep{Wetherill:1991wc,Chambers:2001kt,Obrien:2006jx,Raymond:2006kn,Raymond:2009is,Morishima:2010cs}.  Specifically, simulations that are consistent with the current larger-scale picture of the Solar System's evolution (i.e., including a giant planet instability~\citep{Levison:2011gt}) produce Mars analogs that are an order of magnitude too large~\citep{Obrien:2006jx,Raymond:2009is} (shown as open triangles in panel a of Extended Data Figure 1). Simulations with giant planets on more eccentric orbits can produce improved Mars analogs, however even these improved Mars analogs are often still a factor of two too large at 1.5~AU (also shown as open triangles in panel a of Extended Data Figure 1). Furthermore, classical simulations that produce improved Mars analogs are inconsistent with the late planetesimal-driven migration of the giant planets~\citep{Raymond:2009is,Levison:2011gt} and also tend to form dry terrestrial planets on overly-excited orbits~\citep{Chambers:2002hj,Raymond:2009is}.

These problems served as the main motivation for the Grand Tack scenario, the second category of simulations. Regardless of the initial conditions, which vary as described in the next paragraph, the Grand Tack simulations always proceed as follows. Similar to the classical model, each simulated terrestrial protoplanetary disc starts with a disc of $\sim$~100 embryos and $\sim$~1000--2000 planetesimals. The Grand Tack simulations begin slightly earlier than the classical simulations at 0.6~My before gas dispersal (i.e. $\sim$~2.4~My after the first solids in the Solar System, if the disk disappears after 3~My~\citep{Haisch:2001bx}). \citet{Walsh:2011co} did an extensive exploration of possible giant planet migration evolutions (timescales, initial locations, role of Uranus and Neptune, etc.) and concluded that the sculpting of the inner protoplanetary disc is remarkably insensitive to these details. The key factor is the location of Jupiter's ``tack,'' which occurs when Jupiter and Saturn enter into resonance and change the direction of their migration from inward to outward. Therefore, we use the simplest scheme tested by \citet{Walsh:2011co} namely, ``Saturn's core growing in the 2:3 resonance with Jupiter'' (see Walsh {\it et al.}\cite{Walsh:2011co} supplementary material). In this scheme, during the first 0.1~My, Jupiter and Saturn migrate inward from 3.5 and 4.5~AU to 1.5 and 2~AU, respectively, while Saturn's mass grows linearly from $\sim$~10~M$_\oplus$ to its current mass. This gas-driven migration truncates the embryo and planetesimal disc at $\sim$~1~AU near the 3:2 mean motion resonance with Jupiter. When Saturn reaches a mass close to its final mass, the migration physics changes and Jupiter and Saturn migrate outwards to 5.25 and 7~AU, respectively, in 0.5~My. At this point the gas is removed and all giant planet migration stops. These final locations for Jupiter and Saturn are appropriate initial conditions for the late giant planet instability~\citep{Morbidelli:2007hy}. Then we evolve the simulation for another 150~My. At the end of this period, there are $\sim$~4 terrestrial planets on stable orbits.

For the second category of the simulations, we use the 8 simulations from \citet{Walsh:2011co} and 16 additional simulations from \citet{OBrien:2014bk}, which are statistical variants of the \citet{Walsh:2011co} runs with the addition of outer planetesimals scattered inward by the outward migration of the giant planets. We also use 187 new simulations. The Grand Tack simulations can be grouped into 27 suites. The initial protoplanetary disc conditions are very similar to \citet{Walsh:2011co} with the exception of two parameters: the ratio of the initial amount of mass in embryos to the amount of mass in planetesimals and the initial mass of each individual embryo. We also explored the consequences of an initial distribution of embryos with mass increasing with semi-major axis. 

Regardless of the varied parameters, the mass-orbit distributions are much better than that obtained in the classical simulations (see panel a of Extended Data Figure 1). In panel b, we show the primary result of the main text that the relative late accreted mass is correlated with the time of the last giant impact, and that the Grand Tack scenario gives on average smaller late accreted masses at any given last giant impact time. We note that Grand Tack simulations are consistently more concentrated than the classical simulations independent of the relative late accreted mass (panel c), and are generally as or less dynamically excited than classical simulations (panel d). Nevertheless, those simulations that are consistent with estimates of the late accreted mass (gray horizontal band in panels b-d) produce terrestrial planet systems, which are in general too excited and less concentrated than the real terrestrial planets. This is likely due to the assumption of perfect accretion during collisions, which means that the role of impact ejecta is not accounted for in the simulations. In fact, it has been shown that impact ejecta can significantly reduce the final excitation of the terrestrial planets~\citep{Chambers:2013cp}. This would reduce the angular momentum deficit and increase the concentration parameter, since dynamically colder systems are capable of maintaining a more compact configuration as evidenced by the loose correlation between $S_d$ and $S_c$. Therefore, one should not interpret panels c and d as definitive evidence that the relative late accreted mass should be high, since it's likely that implementing imperfect accretion will change both $S_d$ and $S_c$ metrics for any given late accreted mass. 

Of all the parameters that we changed in the Grand Tack simulations, we found that only the ratio of the total mass of embryos to planetesimals affects the last giant impact time and the late accreted mass (see Extended Data Figure 2). It is clear that by varying the initial disc conditions it is possible to obtain a wide range of last giant impact times and even obtain a wide range of last giant impact times within the same simulation suite. But, regardless of the simulation suite, each last giant impact date corresponds to only a tight range of late accreted masses. 

Naively, it is surprising that the aforementioned range is so tight, since one might think that by changing the initial mass in planetesimals in a simulation, one could get a similar last giant impact time with a different late accreted mass. However, this is not the case because the mass in the planetesimal population governs dynamical friction, which determines when the giant impacts occur. Thus increasing the planetesimal population makes the giant impacts occur earlier and decreasing it makes them occur later. In other words, changing the planetesimal mass makes the final results evolve along a diagonal line (top-left to bottom-right) in Figure 1. This is the reason why for a given last giant impact date, the spread in late accreted mass values is relatively narrow.

The difference between the correlations given by the classical and Grand Tack simulations is due to a fundamental change in the structure of the disk. During the Grand Tack, Jupiter rapidly clears the outer terrestrial region (i.e. asteroid belt and Mars region). Instead, in the classical simulations, this region is de-populated slowly due to resonant interactions with Jupiter. This leads to a slower decay of the planetesimal population and hence a larger mass delivered for a given last giant impact time. We notice however that the dispersions of late accreted mass for a given last giant impact time are comparable for the two scenarios.

\section{Considerations on Late Accretion} In the main text of this paper we showed the existence of a correlation between the timing of the Moon-forming event and the amount of material accreted after this event, known as the late accreted mass. In order to date the Moon-forming event using this correlation, we need a reliable constraint on the late accreted mass. Here we discuss various constraints, which support the usual assumption that the late accreted mass is comparable to the usual estimate of the late veneer mass, i.e. $\sim$~5 $\times$ 10$^{-3}$~M$_\oplus$. Nevertheless, the possibility exists that late accreted mass could have been somewhat larger (but unlikely exceeding 1\% of an Earth mass as we will see below). We proceed in steps below, first presenting a definition of Late Accretion, then introducing a set of compositional and isotopic constraints.

\begin{table}
\begin{center}
\begin{tabular}{@{}llll@{}}
\toprule
& Mantle   & Chondritic  & Mantle to  \\
HSE  & concentration, $C_M$ & concentration, $C_C$ & chondritic ratio, $R$ \\
\midrule
Re & $3.5 \pm 0.6 \times 10^{-10}$ & $5.5 \pm 1.5 \times 10^{-8}$ & $6.3 \pm 2.0 \times 10^{-3}$ \\
Os & $3.9 \pm 0.5 \times 10^{-9}$ & $6.5 \pm 1.8 \times 10^{-7}$ & $6.0 \pm 1.8 \times 10^{-3}$ \\
Ir & $3.5 \pm 0.4 \times 10^{-9}$ & $5.9 \pm 1.9 \times 10^{-7}$ & $5.9 \pm 2.0 \times 10^{-3}$ \\
Ru & $7.0 \pm 0.9 \times 10^{-9}$ & $8.7 \pm 2.5 \times 10^{-7}$ & $8.0 \pm 2.5 \times 10^{-3}$ \\
Pt & $7.6 \pm 1.3 \times 10^{-9}$ & $1.2 \pm 0.3 \times 10^{-6}$ & $6.5 \pm 2.1 \times 10^{-3}$ \\
Pd & $7.1 \pm 1.3 \times 10^{-9}$ & $7.3 \pm 2.2 \times 10^{-7}$ & $9.7 \pm 3.4 \times 10^{-3}$ \\
\bottomrule
\end{tabular}
\end{center}
\end{table}

\begin{table}
\begin{center}
\begin{tabular}{@{}lll@{}}
\toprule
Achondrites &  $\epsilon ^{182}\text{W}_\text{Achon}$  & $\left({\text{Hf}\over\text{W}}\right)_\text{Achon}$   \\
\midrule
Basaltic eucrites & 22 & 27 \\
Aubrites & 11 & 30  \\
Mars & 0.40 & 4.0  \\
\bottomrule
\end{tabular}
\end{center}
\end{table}

\subsection{Introduction to Late Accretion:} After the Moon-forming impact, the Earth continued to grow. In the context of our simulations, we define the late accreted mass $M_{LA}$ as simply the planetesimals accreted by the target Earth-like planet after the last giant impact. A giant impact is a collision between the Earth and a surviving embryo (i.e. oligarch) whereas planetesimals represent the surviving small body population. Embryos are several thousand kilometers in size (i.e. Moon to Mars sized) and planetesimals range up to, possibly, a couple thousand kilometers in size (Ceres, the largest planetesimal remaining in the asteroid belt has a diameter of 900 km). The bimodal mass distribution of embryos and planetesimals is a direct outcome of the oligarchic growth regime of terrestrial planet formation~\citep{Kokubo:1998ka}.

All proposed Moon-forming impacts trigger significant mantle melting and therefore planetary differentiation~\citep{Canup:2008ff,Cuk:2012hj,Canup:2012cd,Reufer:2012dz}. Once the Earth's mantle solidified in the aftermath of the Moon-forming event, the planetesimal bombardment and the remnant radioactive elements were probably not enough of an energy source to re-melt the planet completely. Thus the fate of iron and siderophile elements delivered by late projectiles is determined by the nature of the projectile and the characteristics of its impact. If the largest planetesimals are differentiated, their cores may partially merge with the Earth's core~\citep{Olson:2008hu}. Thus, the termination of terrestrial core formation is ambiguous, i.e. probably not an abrupt or global event. This confuses the link between between Late Accretion and Late Veneer, the latter being a geochemically defined entity, which refers to the mass accreted by the Earth after core closure~\citep{Dauphas:2002hu,Maier:2009kq,Mann:2012fp}.

In order to better constrain the late accreted mass, we assume that all planetesimals, differentiated or not, are bulk chondritic. An undifferentiated planetesimal delivers a mass of chondritic material equal to the planetesimal's mass. For a differentiated planetesimal, a fraction $X$ of the mass of the core may emulsify. In order to explain the excess of HSEs in the Earth relative to the Moon, the emulsified droplets have to be permanently integrated into the Earth's mantle~\citep{Bottke:2010hr}, presumably because the oxidizing conditions in the mantle prevent the iron from reaching the core. The rest $1-X$ of the mass of the projectile's core sinks as a competent blob through the mantle and merges with the Earth's core. The entire mantle of the projectile goes into the Earth's mantle. For convenience, we can split the projectile's mantle into two parts as well with proportions $X$ and $1-X$. In this way, it is clear that a fraction $X$ of the total mass of the planetesimal is delivered as bulk chondritic material to the Earth's mantle. The remaining fraction $1 - X$ of the projectile's mantle provides an achondritic contribution to the Earth's mantle.

This simplification requires us to assume that during Late Accretion no significant mass is delivered by achondritic bodies, i.e. bodies without the metal counterpart that would restore a bulk chondritic composition. These are probably ejecta from the mantles of differentiated embryos or the Earth itself, launched into space during the Moon-forming event~\citep{Reufer:2012dz} or previous hit-and-run and erosive giant impacts~\citep{Asphaug:2006gp,Chambers:2013cp}. This achondritic mass is difficult to constrain using any elemental or isotopic system. Fortunately, we do not need to worry about these fragments. Our simulations address only the chondritic component of late accreted material because the ejection (and potential re-accretion) of mantle material in hit-and-run collisions is not taken into account. This means that we can apply our correlation between the timing of the Moon-forming event and the amount of late accreted material considering solely the mass delivered by bodies of bulk chondritic properties.

\subsection{Highly siderophile element constraints:} 
\label{sec:HSE}
Planetary differentiation within a melted mantle is very efficient~\citep{Rubie:2004fk}, therefore each giant impact including the Moon-forming event, sequesters into the core a large fraction of the highly siderophile elements (HSEs) delivered prior to the impact. Any remaining HSEs in a melted portion of the mantle are strongly fractionated relative to chondritic by the differentiation process as well as being significantly depleted~\citep{Mann:2012fp}. Once the mantle has solidified after a giant impact, it is much more difficult for highly siderophile elements to enter the core unless the mantle is locally melted during the impact of a large planetesimal. In this case, it is possible for the projectile's core to entrain HSEs already resident in the Earth's mantle into the Earth's core. However HSEs in the Earth's mantle are observed to have chondritic proportions to one another~\citep{Becker:2006bi,FischerGodde:2011jd} (see Table~\ref{tab:HSE}). This implies that the vast majority of the HSEs today in the terrestrial mantle did not fractionate during metal-silicate segregation. Reaffirming this hypothesis, the $^{187}$Os/$^{188}$Os isotope ratio (one of the HSEs) is chondritic for both the Earth and the Moon~\citep{Meisel:1996dg,Walker:2004hs,Walker:2009be}.

\begin{table}
\begin{center}
\begin{tabular}{lccc}
\hline \hline
 & Mantle concentration & Chondritic concentration & Mantle to chondritic \\
HSE &  $C_M$ [ng/g] & $C_C$ [ng/g]  & ratio $R$ \\
Re & $0.35 \pm 0.06$ & $55.3 \pm 15.0$ & $6.3 \pm 2.0 \times 10^{-3}$ \\
Os & $3.9 \pm 0.5$ & $653 \pm 180$ & $6.0 \pm 1.8 \times 10^{-3}$ \\
Ir & $3.5 \pm 0.4$ & $592 \pm 185$ & $5.9 \pm 2.0 \times 10^{-3}$ \\
Ru & $7.0 \pm 0.9$ & $872 \pm 253$ & $8.0 \pm 2.5 \times 10^{-3}$ \\
Pt & $7.6 \pm 1.3$ & $1174 \pm 334$ & $6.5 \pm 2.1 \times 10^{-3}$ \\
Pd & $7.1 \pm 1.3$ & $732 \pm 222$ & $9.7 \pm 3.4 \times 10^{-3}$ \\
\hline \hline
\end{tabular}
\caption{\label{tab:HSE} Chondritic concentrations for different HSEs (column 1) in the Earth's mantle~\citep{Becker:2006bi} (column 2) and chondritic meteorites~\citep{Walker:2009be} (column 3) and the ratio of the two (column 4).}
\end{center}
\end{table}

The concentrations and associated 1-$\sigma$ uncertainties of 6 highly siderophile elements~\citep{Becker:2006bi} in the Earth's mantle are reported in Table~\ref{tab:HSE}. We also include the `chondritic mean' and its uncertainty for each of the HSEs, that we determined from the mean, minimum and maximum values for ordinary, carbonaceous and enstatite chondrites reported in \citet{Walker:2009be}. We define the mean chondritic concentrations as the average of the mean values for the three chondrite groups. We estimate the uncertainty by assuming that half the difference between the maximum and the minimum value for each chondrite group is 3-$\sigma$, then average the estimated 1-$\sigma$ values of each chondrite group to obtain the uncertainty of the chondritic concentrations.

The chondritic mass needed to account for the HSE terrestrial mantle budget is $M_{C}= R \times M_{M}$ where $R$ is the ratio of mantle to chondritic abundances of each HSE and $M_{M} = 0.675$ M$_\oplus$ is the mass of the Earth's mantle. Each HSE provides its own chondritic mass estimate, and the 1-$\sigma$ uncertainty is determined by standard error propagation. Averaging them together, we obtain a chondritic mass of $M_{C} = $ 4.8~$\pm$~1.6 $\times$ 10$^{-3}$~M$_\oplus$. 

This chondritic mass corresponds directly to the amount of chondritic material necessary to account for the measured HSE abundances in the Earth's mantle relative to chondritic. We purposely avoid identifying this chondritic mass with either the late veneer mass or the late accreted mass. By definition, the late veneer mass was delivered after core formation ended, and the late accreted mass was delivered after the Moon-forming event. Only under a specific set of assumptions are either of those labels accurate for the chondritic mass, and so we will be careful delineating when different sets of assumptions are being made.

The simplest set of assumptions we can make is to assume that either all late projectiles were undifferentiated or their cores fully dissolved in the Earth's mantle $X=1$ with no metal percolating to the core (presumably because of the oxidized state of the mantle via reaction with~\citep{Frost:2004ck,Frost:2008fb} Fe$^{3+}$ or through the oxidizing effect of water~\citep{Sharp:2013cp}), or some combination of the two. In addition, we assume that the Moon-forming event cleared the entire terrestrial mantle of its previously existing HSEs. Only if all of these assumptions are true, does the late accreted mass $M_{LA}$ coincide with the measured chondritic mass based on HSEs: $$M_{LA} = M_{C} = 4.8 \pm 1.6  \times 10^{-3}~\text{M}_\oplus.$$ In this scenario, no late projectile cores merge with the Earth's core and no mantle HSEs pre-date the Moon-forming impact. This is the assumption most often made in the literature~\citep{Bottke:2010hr,Mann:2012fp,Wang:2013iy} and is the nominal case we present in the main text. Below, we expand this model to a more general case that allows the late accreted mass, the late veneer mass and the chondritic mass to be distinct.

Before we discuss the fate of the projectile cores, we need to assess what portion of Late Accretion was delivered in large bodies with cores. We do this by considering that during Late Accretion, HSEs accumulate not only on the Earth but also on the Moon~\citep{Chyba:1991cq}. However, the lunar chondritic mass~\citep{Day:2007bp,Day:2011uq} is $m_C \approx$ 1.8 $\times$ 10$^{-6}$~M$_\oplus$; it is computed in the same way, from a bulk HSE abundance $\sim$~0.00015 $\times$ chondritic). It should be kept in mind that the concentration value for the Moon is more uncertain than for the Earth because it has been obtained from indirect sampling of the lunar mantle (through igneous rocks such as basalts). The ratio of lunar to Earth chondritic mass is $M_{C} / m_{C} \approx 2700$, whereas the ratio of accreted material according to the ratio of gravitational cross sections is $\sigma \approx 13.5 + 1610 / \left( 5.6 + V_\infty^2 \right)$, where $V_\infty$ is the relative velocity of the projectile before any acceleration from the gravitational potential of the target. The expected flux ratio varies from $\sigma \approx 300$ for bodies with essentially no relative velocity with the Earth-Moon system (i.e. dust), to $\sigma \approx 52$ for asteroids with $V_\infty \approx 6$ km s$^{-1}$ and $\sigma \approx 15$ for comets with $V_\infty \approx 40$ km s$^{-1}$. There is a discrepant factor of 9, 50 or 200 between the measured and the expected ratios of chondritic masses delivered to the Earth and the Moon, depending on the projectile relative velocity $V_\infty =$ 0, 6, or 40~km~s$^{-1}$, respectively.

To reconcile this discrepancy with the assumption that the chondritic mass is entirely delivered during Late Accretion, \citet{Bottke:2010hr} postulated that most of the mass accreted by the Earth was delivered by a few large planetesimals. Since the ratio of gravitational cross-sections only dictate a statistical long-term average of the impact rates, stochasticity and small number statistics can explain why the few largest planetesimals delivered mass only to the Earth and not to the Moon~\citep{Raymond:2013jg}. For instance, only one 2500 km diameter planetesimal is needed to deliver $85\%$ of the chondritic mass assuming a density of $3$~g~cm$^{-3}$. If the largest planetesimals were only 1500 km in diameter, then five such bodies could deliver $93\%$ of the chondritic mass. We quantify this hypothesis by dividing projectiles into two size bins: large $M_L$ and small $M_S$. Thus, $M_{LA} = M_L + M_S$. Large projectiles are rare and thus they are likely to be accreted only by the Earth and not by the Moon, while small projectiles are numerous and therefore they are delivered to the Earth and the Moon proportionately according to their respective cross-sections $\sigma$. \citet{Willbold:2011kw} also finds evidence, using measurements of the Earth's tungsten anomaly, for a `patchy' Late Accretion consistent with a few large impacts rather than many smaller impacts.

The \citet{Bottke:2010hr} scenario is simple and matches the data, but its hypothesis rests principally on two assumptions, both of which are responsible for the conclusion that the late accreted mass is equivalent to the chondritic mass derived from the HSE abundances of the Earth's mantle relative to chondritic meteorites. For a more accurate assessment of the relationship between the late accreted mass and the chondritic mass, we identify each assumption and adjust the model accordingly.

First, we can challenge the assumption that all of the HSEs were delivered late (i.e. after the Moon-forming impact). The survival of mantle HSEs during the Moon-forming event is a possibility to be taken seriously. On the modeling side, new high-resolution smooth-particle hydrodynamic (SPH) simulations of the Moon-forming event show that the opposite hemisphere of the Earth can remain unaffected by the giant collision~\citep{Cuk:2012hj}. While HSEs in the melted hemisphere are transported into the core and are strongly fractionated relative to chondritic, they are also strongly depleted relative to the other hemisphere, where the HSEs retain their chondritic relative proportions. Thus, in this simple scenario, in the end the bulk mantle of the Earth contains HSEs in approximate chondritic relative proportions and depleted roughly by a factor of 2 relative to their pre-Moon-forming-impact abundances. On the geochemical side, \citet{Touboul:2012gq} discuss Kostomuksha komatiites, for which combined $^{182}$W,$^{186,187}$Os, and $^{142,143}$Nd isotopic data indicate that their mantle source underwent metal-silicate fractionation well before 30~My of Solar System history and therefore remained unaffected by the Moon-forming event, assuming that the latter occurred not earlier than this date. This supports the scenario described above. Thus, the HSE budget may be a combination of surviving and late delivered HSEs, and so we quantify the relationship between the chondritic mass $M_C$ corresponding to the mantle budget of HSEs and the chondritic component of the late accreted mass $M_{CLA}$ by introducing a parameter $Y$, which represents the fraction of HSE's delivered during Late Accretion. This parameter can range between $Y = 1$, if nothing pre-dates the Moon-forming event, to $Y = 0$, if everything does. Thus the chondritic component of the late accreted mass is related to the chondritic mass determined from the HSEs: $M_{CLA} = Y M_C$.

Second, we can challenge the assumption that that the projectile cores merged entirely with the Earth's mantle. This requires that the core is emulsified or at least broken into incoherent pieces that interact efficiently with the mantle~\citep{Rubie:2003hq} and oxidized via reaction with Fe$^{3+}$~\citep{Frost:2004ck,Frost:2008fb,Bottke:2010hr} or alternatively with water~\citep{Sharp:2013cp} to prevent the transportation of iron and siderophile elements into the Earth's core. Recent results suggest that it is possible for part of the projectile's core to remain intact and merge directly with the Earth's core. Using smooth-particle hydrodynamics simulations, Robin Canup ({\it personal communication}) finds that only up to half of the largest projectile cores can merge with the Earth's core. On the other hand, physical experiments~\citep{Olson:2008hu,Deguen:2011ig} and numerical simulations~\citep{Dahl:2010ik,Samuel:2012gw} of metal sinking through the mantle conclude that projectile cores with radii much smaller than the depth of the magma ocean emulsify efficiently. From these studies, it is unclear whether larger planetesimal cores may survive the descent towards the core-mantle boundary.  As we already said, we parameterize this uncertainty by asserting that an average fraction $X$ of each core dissolves in the mantle of the Earth, while the remaining fraction $1-X$  merges into the Earth's core. The $X$ parameter can range between 1 (complete assimilation of the projectile's core into the Earth's mantle) and 0 (the entire projectile's core sinking onto the Earth core). We stress that the fraction $X$ represents the mass weighted average for all the late accreted large planetesimals, not each individual one:$$X = \frac{1}{M_L} \sum_{i}^{N} M_i X_i,$$ where $N$ is the total number of large planetesimals, $M_i$ and $X_i$ are their individual masses and values of $X$. In this average, if large undifferentiated planetesimals, for instance very oxidized bodies, contribute to Late Accretion, then they count as if they have $X_i=1$.

We had already divided the late accreted mass into components that were delivered either by small and large bodies: $M_{LA} = M_L + M_S$. Reminding ourselves that we have assumed that all bodies are chondritic in bulk, small bodies thus contribute their entire mass to the chondritic component of the late accreted mass. Large and hence differentiated bodies only deliver on average a fraction $X$ of their mass to the chondritic component of the late accreted mass. Thus, the chondritic component of the late accreted mass is: $$M_{CLA} = X M_L + M_S = Y M_C.$$ Using this set of relationships and the expression of the late accreted mass as the sum of two size components: $M_{LA} = M_L + M_S$, we can re-express the late accreted mass as a function of the unknown parameters $X$ and $Y$: 
$$
M_{LA} = \frac{1}{X} (Y M_C - \sigma m_C ) + \sigma m_C = M_C \frac{Y}{X} \left( 1 - \sigma \frac{ m_C}{ M_C} \left( \frac{ 1 - X }{Y} \right) \right),
$$ 
where $M_C = 4.8 \times 10^{-3}$ is the chondritic mass in the Earth's mantle, $m_C = 1.8 \times 10^{-6} M_\oplus$ is the lunar late accreted mass, $\sigma$ is the cross-section ratio between the Earth and the Moon, and $M_S=\sigma m_C$ is the mass delivered by small projectiles to the Earth. Because $\sigma$ is much smaller than the ratio $M_C/m_C$ the following analysis is not sensitive to the choice of $\sigma$. 

Extended Data Figure 3 shows the late accreted mass $M_{LA}$ as a function of $X$ and $Y$ given the terrestrial $M_C$ and lunar chondritic masses $m_C$, and a gravitational cross-section ratio $\sigma = 52$ appropriate for planetesimals with semi-major axes in the terrestrial region. As the parameters $X$ and $Y$ change, the mass delivered in large planetesimals must adjust to continue to match the HSE constraints. Immediately, we see that a late accreted mass of 5 $\times$ 10$^{-3}$~M$_\oplus$ approximately divides the parameter space in half, running diagonally from near the \citet{Bottke:2010hr} scenario, which is exactly at $X = 1$ and $Y = 1$, to a scenario where all of the HSEs pre-date the Moon-forming impact and the core of each projectile accretes entirely into the Earth's core $X=Y=0$. Scenarios below this contour in the figure require smaller late accreted masses than the \citet{Bottke:2010hr} scenario, while those above require more. Most of the parameter space has a late accreted mass $M_{LA} <$ 0.01~M$_\oplus$.

\subsection{Constraining X:} Up to now, $X$ has been treated as a free parameter; as $X$ approaches zero, the late accreted mass can be arbitrarily large. We now try to constrain $X$ by addressing what fraction of the projectile's core emulsifies after impact with the Earth. The emulsification of the projectile's core is a mechanical (hydrodynamical) process that does not depend on the chemical properties of the Earth, which instead determine the fate of the metallic droplets. For this reason, we can use works that constrain the fraction of the projectile emulsified in an impact~\citep{Rudge:2010fv,Dahl:2010ik,Deguen:2011ig,Samuel:2012gw} regardless of what these work envision happening to the droplets. The droplets can percolate into the Earth's core if the mantle is sufficiently reduced as in the early phases of accretion~\citep{Rubie:2011cr} or be oxidized and remain in the mantle as necessitated to explain the HSE abundances and their chondritic relative proportions in the Earth's mantle~\citep{Bottke:2010hr}. Physical experiments~\citep{Olson:2008hu,Deguen:2011ig} and numerical simulations~\citep{Dahl:2010ik,Samuel:2012gw} show that emulsification is important until the projectile core approaches the thickness of the mantle; however, it is difficult to derive a lower bound on $X$ given the problems in modeling the emulsification process which involves length scales ranging from centimeters to hundreds of kilometers or more. Thus, we think that the best constraints on $X$ can be derived from geochemical analysis~\citep{Rudge:2010fv}.

Specifically, we apply the model of \citet{Rudge:2010fv} for the evolution of the mantle's tungsten isotopic anomaly during the accretion history. The coefficient $k$ in \citet{Rudge:2010fv} determines the fraction of the projectile cores that emulsify in the Earth's mantle like our coefficient $X$, so they have the same meaning for our purposes.
 
\citet{Rudge:2010fv} showed that isotopic constraints from the Hf-W and U-Pb systems bound the accretion history of the Earth. If growth is too fast or too slow, too much or too little radiogenic tungsten and lead accumulate in the Earth's mantle. The bounds presented in \citet{Rudge:2010fv} assume constant partitioning coefficients, which in reality are a function of temperature, pressure and oxygen fugacity. The oxygen fugacity can change significantly during the early accretion history~\citep{Wade:2005fi,Rubie:2011cr} while the temperature and pressure are required to lie on the peridotite liquidus. However, \citet{Rudge:2010fv} discovered that isotopic evolution models with and without evolving partition coefficients do not differ substantially. The only difference they found when using evolving partition coefficients ``is that accretion needs to be slightly more protracted to match the same Hf/W and U/Pb observations. The requirement of more protracted accretion arises because both W and Pb are more siderophile during the early accretion than the Late Accretion, which causes a bias towards younger ages.'' 

Panel a in Extended Data Figure 4 shows the $X \geq 0.53$ bound determined following \citet{Rudge:2010fv} from Hf-W constraints (U-Pb does not constrain the equilibration factor). This bound is not identical to that stated in \citet{Rudge:2010fv} (0.36), since we have used the updated terrestrial Hf/W measurements of \citet{Konig:2011we}. Examining panel a, we see that all combinations of $X \gtrsim 0.53$ and $Y$ produce late accreted masses $M_{LA} \leq 8.9 \times 10^{-3}$ $M_\oplus$.

\subsection{Isotopic Constraints} Additional constraints on the late accreted mass of the Earth come from the fact that the Earth and the Moon have very similar isotopic contributions. If a large mass of a specific composition had been added to the Earth after the Moon forming event, then the Earth and the Moon would be more different than observed. Here we first examine the isotopic systems of oxygen and titanium, which show that the terrestrial late accreted mass is unlikely to exceed 1\% of an Earth mass regardless of the dominant projectile composition with the possible exception of enstatite chondrites. A differentiated enstatite chondrite however should have a strong radiogenic tungsten signature given its reduced nature. Therefore, we next examine constraints provided by the lunar-terrestrial tungsten anomaly, which is the difference between the tungsten isotopic composition of the Moon and Earth.

\noindent {\it Oxygen and titanium isotopic systems:} If most of the late accreted mass was delivered by one or a few large differentiated planetesimals, as required to explain the difference in HSE abundances of the Earth and Moon, its chemical nature could be easily dominated by a single parent body composition. Different isotopic systems can constrain contributions to the late accreted mass for projectile compositions analog to most meteorite types.

\citet{Wiechert:2001il} and \citet{Spicuzza:2007cb} found that the Earth and the Moon lay on the same oxygen isotope fractionation line $\Delta^{17}$O$_\text{Moon} - \Delta^{17}$O$_\text{Earth} = 0.008 \pm 0.01 \permil$ where $\Delta ^{17}$O$_\text{Sample} = \delta ^{17}$O$_\text{Sample} - 0.5245 \times \delta ^{18}$O$_\text{Sample}$ and $\delta ^{i}$O$_\text{Sample} = 10^3 \times \big[ ( ^{i}$O$ / ^{16}$O$ )_\text{Sample}  / ( ^{i}$O$ / ^{16}$O$ )_\text{Standard} - 1 \big] $. All reported uncertainties are 1-$\sigma$. Recently, Herwartz and colleagues announced at the Royal Society/Kavli Institute meeting on the {\it Origin of the Moon -- challenges and prospects} (25-26 September, 2013) at Chicheley Hall, Buckinghamshire, U.K. that they resolve a difference between the two bodies: $\Delta^{17}$O$_\text{Moon} - \Delta^{17}$O$_\text{Earth}  = 0.012 \permil$. If this measurement is correct and if we assume that the Earth and Moon equilibrated oxygen isotopes at the time of the Moon-forming event,  knowing that the Earth received more late accreted mass than the Moon, we conclude that the dominant nature of the projectiles during Late Accretion must have had a carbonaceous chondrite (excluding CI), enstatite or HED composition, because these are the only meteoritic compositions that are situated below or on the terrestrial fractionation line in the oxygen-isotope diagram. 

Then we can constrain how much late accreted mass could be added to the Earth from carbonaceous CV, CO, CK, CM, CR and CH meteoritic compositions using their measured oxygen isotope compositions~\citep{Clayton:1999hm}: $\Delta^{17}$O$_\text{CV} = - 3.3 \pm 0.6 \permil$, $\Delta^{17}$O$_\text{CO} = - 4.3 \pm 0.3 \permil$, $\Delta^{17}$O$_\text{CK}= - 4.2 \pm 0.4 \permil$, $\Delta^{17}$O$_\text{CM}= - 2.3 \pm 1.1 \permil$, $\Delta^{17}$O$_\text{CR}= - 1.6 \pm 0.5 \permil$ and $\Delta^{17}$O$_\text{CH}= - 1.5 \pm 0.2 \permil$. A late accreted mass dominated by each composition is limited to be less than $0.4 \pm 0.3\%$, $0.3 \pm 0.2\%$, $0.3 \pm 0.2\%$, $0.5 \pm 0.5\%$, $0.8 \pm 0.7\%$ and $0.8 \pm 0.7\%$ of an M$_\oplus$, respectively. For HEDs and enstatites, oxygen isotopes are not very useful, because these meteorite groups are much closer to the terrestrial fractionation line. 

Further constraints are provided by the relative titanium isotope composition of the Earth and Moon. The Moon-Earth difference is~\citep{Zhang:2012dn} $\epsilon ^{50}$Ti$_\text{Moon} - \epsilon ^{50}$Ti$_\text{Earth} = - 0.04 \pm 0.02$ where $\epsilon ^{50}$Ti$_\text{Sample} = 10^4 \times \big[ ( ^{50}$Ti$ / ^{47}$Ti$ )_\text{Sample} / ( ^{50}$Ti$ / ^{47}$Ti$ )_\text{Standard} - 1 ]$. HEDs and, to a lesser extent, enstatites, can be excluded as the dominant component of Late Accretion because they have negative titanium isotope compositions~\citep{Zhang:2012dn} ($\epsilon ^{50}$Ti$_\text{Enstatite} = -0.23 \pm 0.09$ and $\epsilon ^{50}$Ti$_\text{HED} = -1.24 \pm 0.03$) and any large addition of this composition would result in a positive $\epsilon ^{50}$Ti measurement difference between the Moon and the Earth. A source of uncertainty in this analysis appears from the possibility that the Moon-Earth difference is positive, since the 2-$\sigma$ uncertainty encompasses the zero value. Considering this possibility and assuming that $\epsilon ^{50}$Ti$_\text{Moon} - \epsilon ^{50}$Ti$_\text{Earth} = 0.01$ (a 2.5-$\sigma$ deviation), we can still constrain how much late accreted mass could be added to the Earth from HED meteoritic compositions to $0.8 \%$ of an M$_\oplus$, however we cannot place such a strict constraint on an enstatite chondrite composition.

In summary, oxygen and titanium isotopic constraints limit the dominant component of the late accreted mass to less than 1\% of an Earth mass for most meteoritic compositions. The only exception may be an enstatite chondritic composition, if the Earth-Moon difference is $\epsilon ^{50}$Ti$_\text{Moon} - \epsilon ^{50}$Ti$_\text{Earth} \geq 0.002$, which is barely exterior to the upper 2-$\sigma$ uncertainty of the current measurements, then a considerable mass with enstatite composition could be added to the Earth. However, we can probably still exclude differentiated enstatite chondritic bodies from dominating Late Accretion. The mantle's isotopic compositions of ruthenium (delivered during Late Accretion) and molybdenum (delivered throughout accretion) lie along the meteoritic mixing line~\citep{Dauphas:2004hx}. This suggests that the isotopic nature of the projectiles did not change significantly before and after the Moon-forming impact. Given that \citet{Fitoussi:2012kp} concluded, on the basis of silicon isotope compositions, that the enstatite chondrite contribution to the bulk Earth is $< 15\%$ using silicon isotope ratios, it is unlikely that the late accreted mass was dominated by enstatite chondrite projectiles. Moreover, \citet{Wang:2013iy} also show a limited contribution of ordinary and enstatite chondrites to Late Accretion from an analysis of the Earth's Se/Ir and Te/Ir ratios.

As stated at the beginning of this section, the number of bodies delivering most of Late Accretion must be small. For a single isotopic system, it is possible that the isotopic anomalies of the few largest projectiles relative to the Earth-Moon system cancel out. However, it is difficult for this to happen for two or more isotopic systems because the anomalies are different. For instance $\Delta$O$^{17}$ for ordinary chondrites and CR chondrites are opposite in sign and almost the same in absolute magnitude, but the $\epsilon$Ti$^{50}$ of CR chondrites is five times bigger than that of ordinary chondrites.

\noindent {\it Lunar-terrestrial tungsten anomaly:} Given that the Earth and the Moon have identical (at analytic precision) isotope compositions for many non-radiogenic elements (O~\citep{Wiechert:2001il,Spicuzza:2007cb}, Cr~\citep{Lugmair:1998iq,Trinquier:2008bp}, Ti~\citep{Leya:2008ik} and Si~\citep{Armytage:2012br,Fitoussi:2012kp}), it is reasonable to assume that the Earth and the Moon started off with identical W compositions as well. Today, the difference in $^{182}$W isotope composition between the Moon and the Earth is~\citep{Touboul:2007is} $$\epsilon^{182}\text{W}_\text{Moon} - \epsilon^{182}\text{W}_\text{Earth}  = 0.09 \pm 0.1,$$ where $\epsilon^{182}$W$_\text{Moon}$ and $\epsilon^{182}$W$_\text{Earth}$ are contributions to the tungsten anomaly after the Moon-forming impact until now, and $\epsilon^{182}$W$_\text{Sample} = 10^4 \times \big[ ( ^{182}$W$ / ^{184}$W$ )_\text{Sample} / ( ^{182}$W$ / ^{184}$W$ )_\text{Standard} ]$. 

Each contribution can be broken into two categories: $\epsilon^{182}$W$_\text{Sample} = \epsilon^{182}$W$_\text{Sample}^\text{Small} + \epsilon^{182}$W$_\text{Sample}^\text{Large}$, contributions from small and large bodies, respectively. We assume small, proportionately delivered bodies provide material of bulk chondritic composition. The small projectile contribution to the lunar tungsten anomaly is:
$$
\epsilon^{182}\text{W}_\text{Moon}^\text{Small} = \epsilon^{182}\text{W}_\text{Chon} \frac{\left({\text{Hf}\over\text{W}}\right)_\text{Moon}}{\left({\text{Hf}\over\text{W}}\right)_\text{Chon}} \left( \frac{m_C}{m_m} \right) = - 4333 \frac{m_C}{\text{M}_\oplus},
$$
where $\epsilon^{182}\text{W}_\text{Chon} = -2$ is the average of the $^{182}$W abundance of chondrites~\citep{Kleine:2009tp}, $\left({\text{Hf}\over\text{W}}\right)_\text{Moon} = 26$ and $\left({\text{Hf}\over\text{W}}\right)_\text{Chon} = 1$ are the halfnium/tungsten ratios of the Moon~\citep{Kleine:2009tp} and the chondritic average~\citep{Kleine:2009tp}, respectively, $m_C = 1.8 \times 10^{-6} M_\oplus$ is the lunar late accreted mass~\citep{Day:2007bp,Day:2011uq} which is delivered entirely by small projectiles, and $m_m = 0.012$ M$_\oplus$ is the mass of the Moon's mantle. 

For the Earth, the small projectile contribution to the terrestrial tungsten anomaly is:
$$
\epsilon^{182}\text{W}_\text{Earth}^\text{Small}= \epsilon^{182}\text{W}_\text{Chon} \frac{\left({\text{Hf}\over\text{W}}\right)_\text{Earth}}{\left({\text{Hf}\over\text{W}}\right)_\text{Chon}} \left( \frac{M_S}{M_M} \right) = - 77 \frac{\sigma m_C}{\text{M}_\oplus},
$$
where $\left({\text{Hf}\over\text{W}}\right)_\text{Earth} = 26$ is also the Earth's hafnium/tungsten ratio~\citep{Konig:2011we}, $M_S = \sigma m_C$ is the mass of small projectiles impacting the Earth in terms of the lunar late accreted mass and the relative gravitational cross-section of the Earth and Moon since small projectiles are delivered proportionately, and $M_M = 0.675$ M$_\oplus$ is the mass of the Earth's mantle.

Large bodies are stochastically delivered according to \citet{Bottke:2010hr} and thus presumably contributed only to the Earth. Thus they do not contribute to the Moon $\epsilon ^{182}$W$_\text{Moon}^\text{Large} = 0$. Earth contributions are complicated by the different possible fates of the projectile's core, which is captured by the parameter $X$. If the projectile's core is completely accreted in the Earth's mantle $X=1$, then the planetesimal contributes like a chondritic body;  however if $X<1$ then the tungsten anomaly contribution has two components: $\epsilon ^{182}$W$_\text{Earth}^\text{Large} = \epsilon ^{182}$W$_\text{Earth}^\text{Chon} + \epsilon ^{182}$W$_\text{Earth}^\text{Achon}$. The first component corresponds to the fraction $X$ of the projectile that merges with the Earth's mantle delivering a chondritic contribution:
$$
\epsilon ^{182}\text{W}_\text{Earth}^\text{Chon}  = \epsilon ^{182}\text{W}_\text{Chon} \frac{\left({\text{Hf}\over\text{W}}\right)_\text{Earth}}{\left({\text{Hf}\over\text{W}}\right)_\text{Chon}} \left( \frac{X M_{L}}{M_M} \right) =  - 77 \frac{Y M_C - \sigma m_C }{\text{M}_\oplus},
$$
where we used the relationship $X M_L=Y M_C - \sigma m_c$.

The other component corresponds to the $1-X$ fraction of each projectile's core that sinks and merges with the Earth's core. This component contributes only projectile mantle (achondritic) material to the terrestrial mantle tungsten anomaly:
\begin{align}
\epsilon ^{182}\text{W}_\text{Earth}^\text{Achon} & =   \epsilon ^{182}\text{W}_\text{Achon} \frac{\left({\text{Hf}\over\text{W}}\right)_\text{Earth}}{\left({\text{Hf}\over\text{W}}\right)_\text{Achon}} \left(\frac{0.675 (1-X) M_{L}}{M_M} \right) \\ & =  \frac{ \epsilon ^{182}\text{W}_\text{Achon}}{\left({\text{Hf}\over\text{W}}\right)_\text{Achon} } \times 26 \left( \frac{ \left( 1 - X  \right) \left( Y M_C - \sigma m_C \right)  }{ X \text{M}_\oplus} \right),
\end{align}
where we have assumed, for simplicity that the fraction of the projectile mantle to its total mass and the Earth's mantle  to its total mass are the same, i.e. 0.675. The formula above contains two unknown parameters for the mantle material of the large projectiles: $\epsilon ^{182}$W$_\text{Earth}^\text{Achon}$ and $\left({\text{Hf}\over\text{W}}\right)_\text{Achon}$. Good analogs for the mantles of differentiated bodies come from the achondrite meteorite collections, and we list three examples in Table~\ref{tab:achond}: basaltic eucrites, aubrites and martian meteorites:

\begin{table}[h]
\begin{center}
\begin{tabular}{lcc}
\hline \hline
Achondrites &  $\epsilon ^{182}\text{W}_\text{Achon}$  & $\left({\text{Hf}\over\text{W}}\right)_\text{Achon}$   \\
Basaltic Eucrites~\citep{Kleine:2009tp} & 22 & 27 \\
Aubrites~\citep{Petitat:2008uj} & 11 & 30  \\
Mars~\citep{Kleine:2009tp} & 0.4 & 4.0  \\
\hline \hline
\end{tabular}
\caption{\label{tab:achond} Tungsten abundances $\epsilon ^{182}\text{W}_\text{Achon}$ (column 2) and $\left({\text{Hf}\over\text{W}}\right)_\text{Achon}$ ratios (column 3) for achondritic mantle materials (column 1).}
\end{center}
\end{table}

For each analog, we calculate the terrestrial tungsten anomaly contribution from differentiated mantles $\epsilon ^{182}\text{W}_\text{Earth}^\text{Achon}$ and then the tungsten difference between the Moon and the Earth $\epsilon^{182}\text{W}_\text{Moon} - \epsilon^{182}\text{W}_\text{Earth} $ from the expressions above. This expression must match  $0.09 \pm 0.1$ and is a function of $X$ and $Y$. We overlay this tungsten difference constraint on top of the parameter space map shown in Panel a of Extended Data Figure 4 to create new maps (see panels b-d) showing the contour that matches the nominal tungsten difference of 0.09 as a green solid line, as well as the 1-$\sigma$ limits as green dashed lines (regions exterior to the 1-$\sigma$ uncertainty are red).

Regardless of the analog, the lunar-terrestrial tungsten difference places an upper and lower limit to the late accreted mass. It also always removes the possibility that the simple assumptions made in \citet{Bottke:2010hr} are correct since the lunar-terrestrial tungsten difference is always too large when $X=1$ and $Y=1$. \citet{Bottke:2010hr} checked this constraint at the time of writing, but we are using here the updated \citet{Konig:2011we} values for $\left({\text{Hf}\over\text{W}}\right)_\text{Earth} = 26$. 

Using eucrite and the aubrite compositions rules out late accreted masses exceeding $0.013$ and $0.033$ M$_\oplus$, respectively, but most of the parameter space consistent with the tungsten difference constraint implies a late accreted mass $M_{LA} \lesssim 0.01$ M$_\oplus$, particularly when considering eucrite-like impactors. Only for a Mars-like composition is the constraint looser. We stress that enstatite chondrites are very reduced, so a large enstatite body which underwent an early differentiation is expected to have a large hafnium-tungsten ratio and a large tungsten isotopic anomaly more similar to what is observed in the HEDs and aubrites.

\subsection{Conclusions on Late Accretion:} The Earth mantle HSE budget, mantle tungsten isotope composition, the small difference in tungsten isotope composition between the Moon and the Earth and other isotopic systems constrain the late accreted mass $M_{LA}$. Assumptions need to be made on the nature of the late accreted projectiles, but it turns out to be unlikely that the late accreted mass was larger than $0.01$ M$_\oplus$. The standard model, which assumes that the late accreted mass is the same as the chondritic mass ($M_{LA} = M_C = 4.8 \pm 1.6 \times 10^{-3}$), seems quite correct. Moreover it is also consistent with other non-geochemical constraints. In fact, Simone Marchi ({\it personal communication}) reports that in his Monte Carlo simulations of the \citet{Bottke:2010hr} scenario, calibrated on the lunar impact flux, the total mass delivered to the Earth results in less than $1\%$ of the Earth's mass in the vast majority of cases. Scenarios delivering more than $0.01$ M$_\oplus$ in late accreted mass, while being potentially consistent with the constraints discussed in this paper, require ad-hoc assumptions on the chemical and isotopic properties of the dominant projectiles.

\begin{figure}
\center
\includegraphics[width=0.9\textwidth]{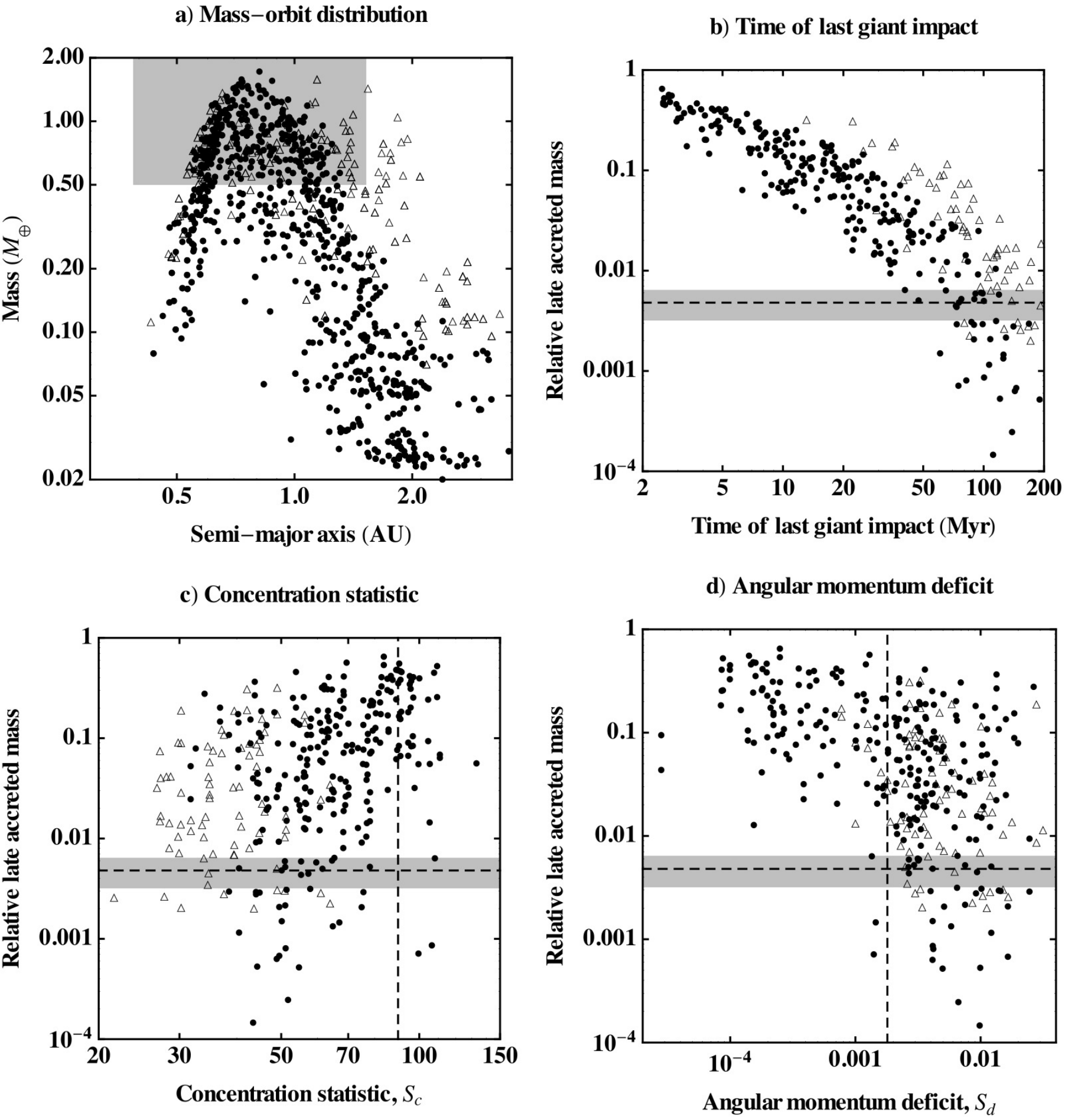}
\caption{Four panels comparing the classical (triangles) and Grand Tack (circles) scenarios. Panel a shows the mass of each planet in every simulated Solar System as a function of semi-major axis. Planets within grey zone are considered Earth-like. Planets at Mars' distance from the Sun are too massive in classical simulations but are correct in Grand Tack simulations. The next three panels show the relative Late Accretion mass as a function of different parameters and the chondritic mass and uncertainty $M_C = $4.8 $\pm$ 1.6 $\times$ 10$^{-3}$ M$_\oplus$ as a dashed line and grey region. Panel b is the same as Figure 1 of the main text, but with a broader time scale. Panel c and d show the relative Late Accretion mass and two orbital structure statistics: concentration $S_c$ and angular momentum deficit $S_d$. The concentration statistic is defined~\citep{Chambers:2001kt}: $ S_c = \max \left( \frac{\sum_j m_j }{\sum_j m_j \left( Log_{10} \left(a /a_j\right) \right)^2 } \right),$ where the summation is over the planets and $m_j$ and $a_j$ are the mass and semi-major axis of the $j$th planet. $S_c$ is then the maximum of this function as $a$ is varied. The Solar System has $S_c = 89.9$ and in panel c is marked with a vertical dashed line. Grand Tack simulations are as or more concentrated than classical simulations. The angular momentum deficit statistic is defined~\citep{Laskar:1997vw,Chambers:2001kt}: $ S_d = \frac{\sum_j m_j \sqrt{a_j} \left( 1- \sqrt{1-e_j^2 } \cos i_j \right)}{\sum_j m_j \sqrt{a_j}},$ where the summation is again over the planets and $e_j$ and $i_j$ are the eccentricity and inclination of the $j$th planet. The Solar System has $S_d = 0.0018$ and this marked by a vertical dashed line in panel d. A late giant planet instability ($\sim$500~My after first Solids in the Solar System) likely excites the inner Solar System, since these simulations end before the instability they should only be a fraction of the current value~\citep{Brasser:2013hs}. Grand Tack simulations are either as dynamically excited or colder than classical simulations, although they are in some cases hotter than the current terrestrial planets.}
\end{figure}

\begin{figure}
\center
\includegraphics[width=\textwidth]{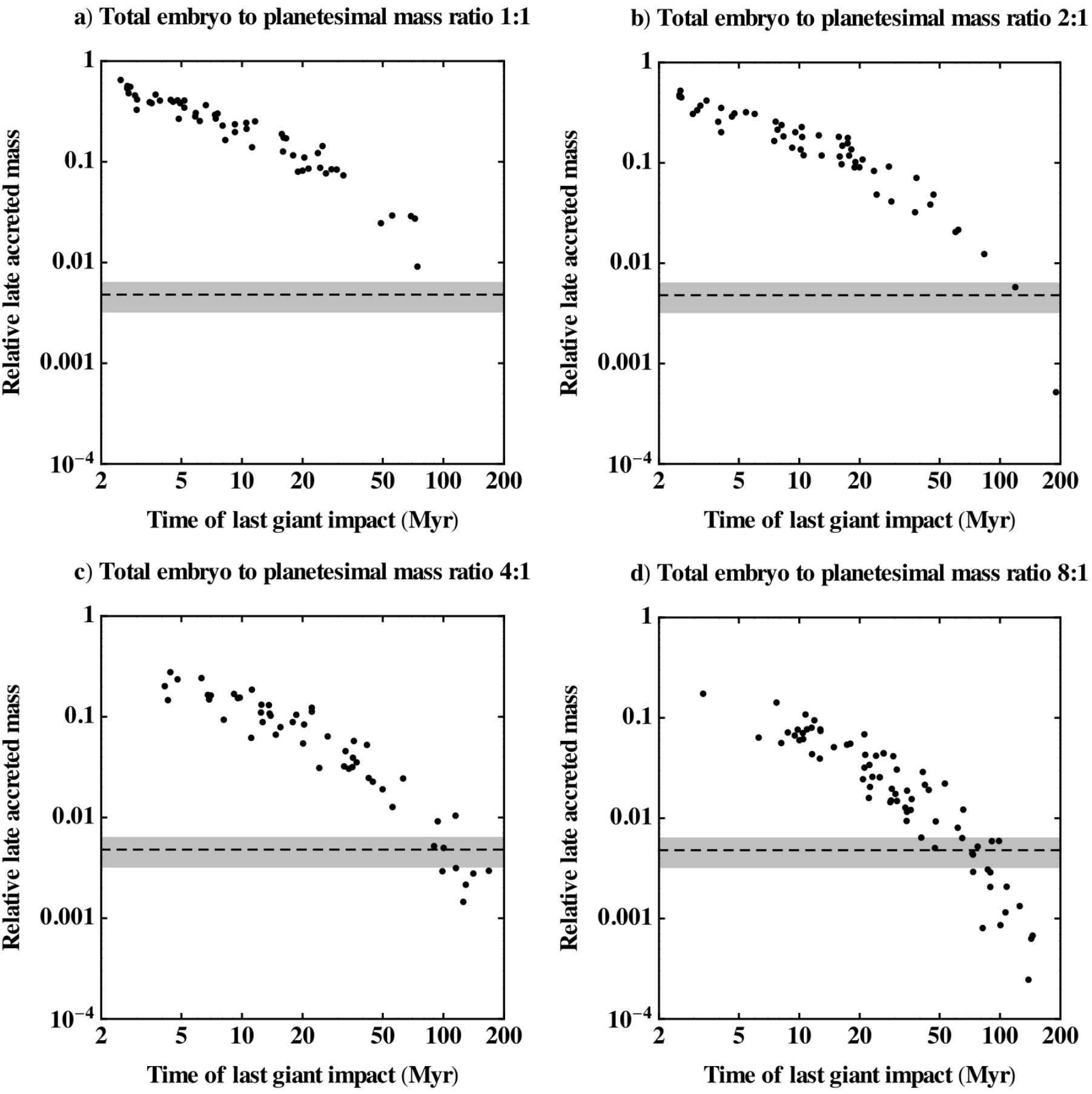}
\caption{Figure 1 is reproduced in each panel, showing the Earth-like planet relative late accreted mass as a function of last giant impact time for each scenario within the classical category. For reference, the horizontal dashed line and enclosing darkened region are the best estimate and 1-$\sigma$ uncertainty of the late veneer mass $M_{C}$ = 4.8 $\pm$ 1.6$\times$10$^{-3}$ M$_\oplus$. Each panel shows the Earth-like planet relative late accreted mass as a function of last giant impact time for different initial ratios of the total mass in embryos to planetesimals including 1/1 in a, 2/1 in b, 4/1 in c, and 8/1 in d. The mean time of the last giant impact increases and the relative Late Accretion mass decreases as the initial total embryo to planetesimal mass ratio increases.}
\end{figure}

\begin{figure}
\centering
\includegraphics{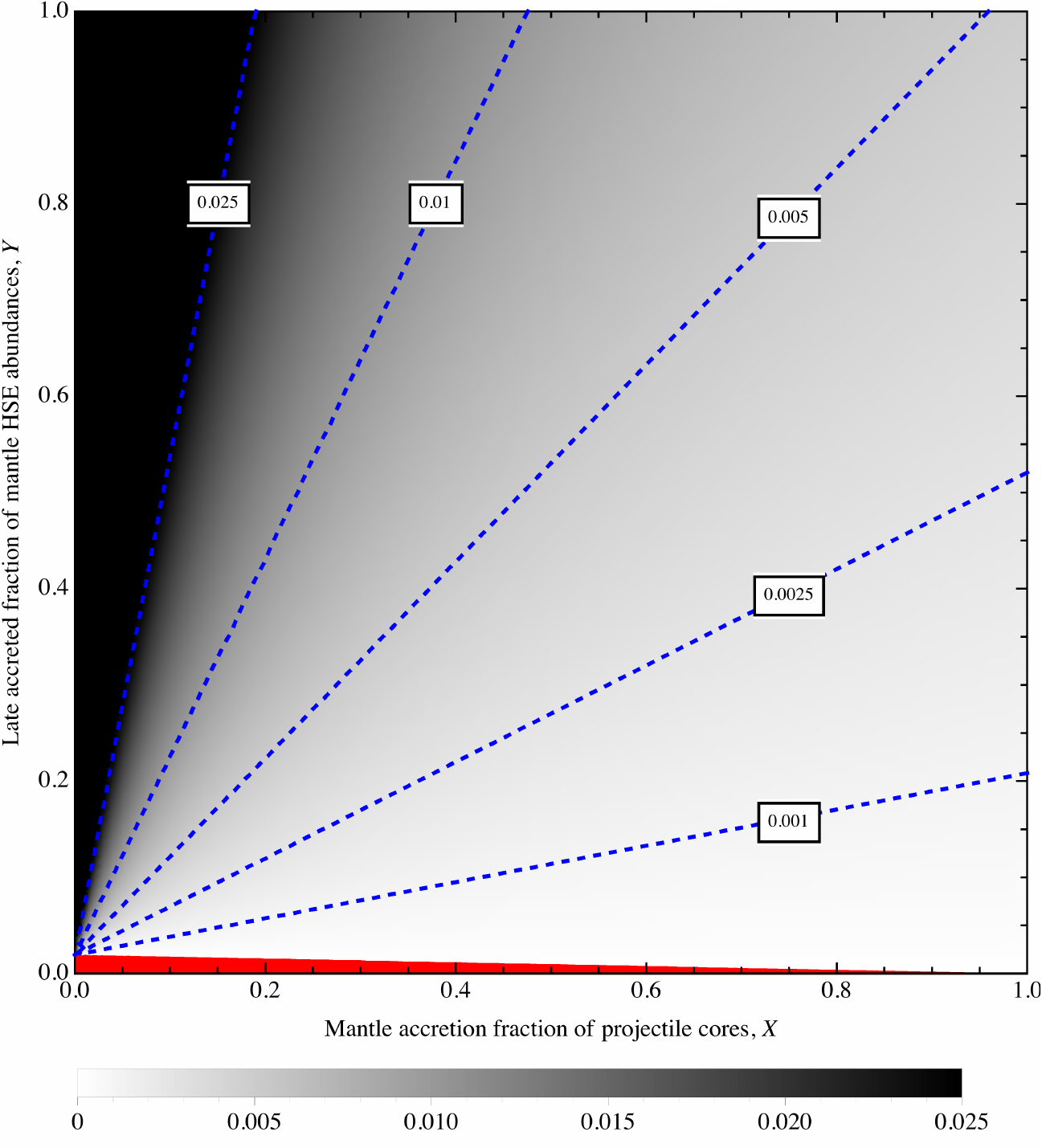}
\caption{The late accreted mass $M_{LA}$ shown as grayscale contours of the two parameters: $X$ and $Y$. The blue dashed contours indicate the location of a few specific late accreted masses and are labeled in the framed boxes. For this figure, we assumed that the Earth receives 52 times as much mass in small projectiles than the Moon ($\sigma = 52$). The red region is inaccessible, since the measured chondritic mass in the lunar mantle requires a minimum flux of small planetesimals onto the Earth's mantle.}
\end{figure}

\begin{figure}
\centering
\includegraphics{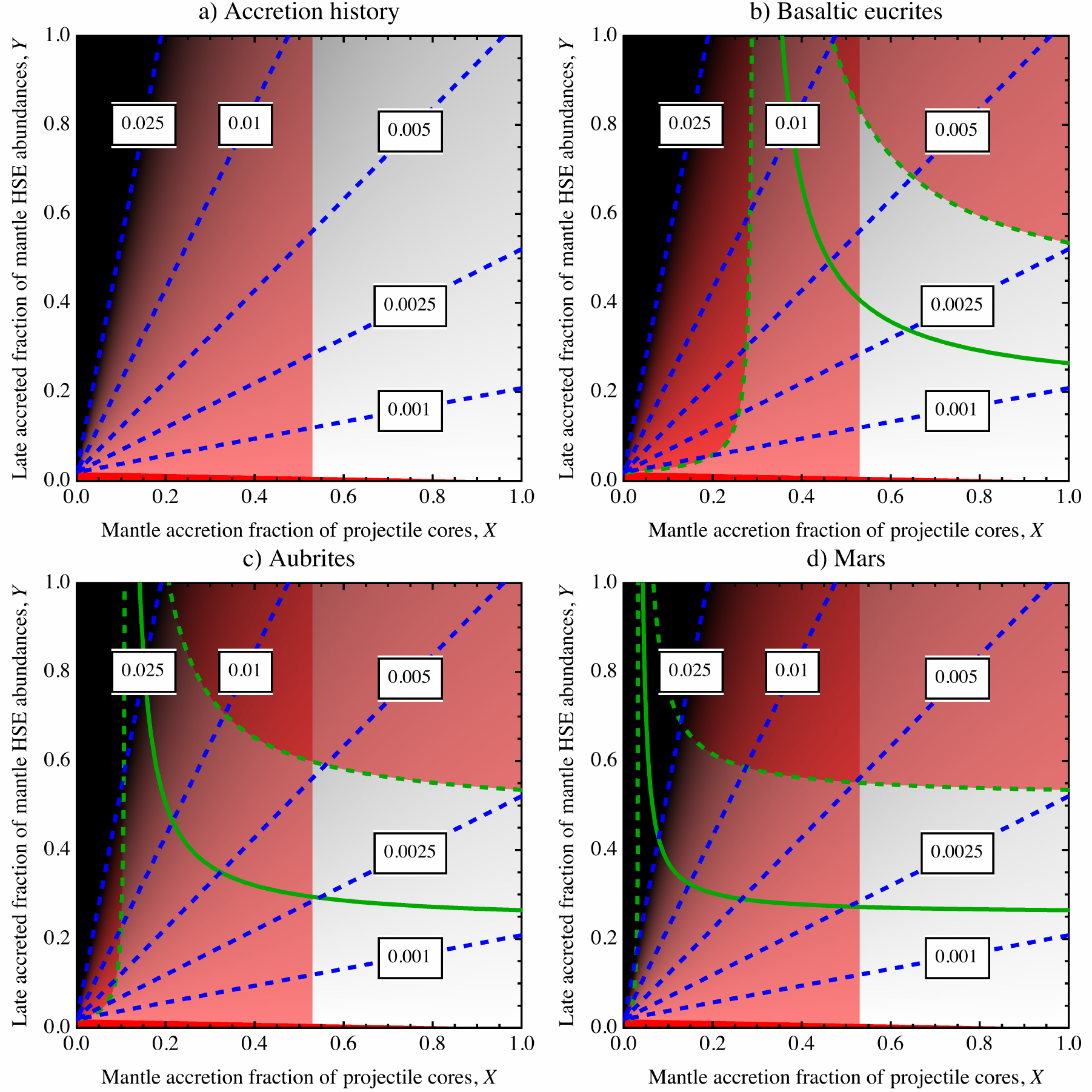}
\caption{Each panel is like Extended Data Figure 3 but also showing additional constraints. The constraint $X \geq 0.53$, deduced by a reanalysis of \citet{Rudge:2010fv} with the Hf/W ratio updated from \citet{Konig:2011we}, is shown in light red in each panel. In panels b-d, the solid green contours indicate the set of $X,Y$ parameters that reproduce the nominal difference in W-isotope composition between the Moon and the Earth (0.09 in $\epsilon$ units; see \citet{Touboul:2007is}). The difference among these panels is in the assumed Hf/W and W-isotope composition for the mantle of the differentiated projectiles: panel b assumes values typical of eucrites; panel c is for aubrites and panel d is for Mars.
The green dashed contours are the 1-$\sigma$ uncertainties on $X,Y$ related to the uncertainty on the difference between the W-isotope composition between the Moon and the earth: $\pm 0.1$ in $\epsilon$ units. The areas exterior to these $1-\sigma$ uncertainties have been colored red.}
\end{figure}

\clearpage

\bibliography{biblio}

\begin{thebibliography}{78}
\providecommand{\natexlab}[1]{#1}
\expandafter\ifx\csname urlstyle\endcsname\relax
  \providecommand{\doi}[1]{doi:\discretionary{}{}{}#1}\else
  \providecommand{\doi}{doi:\discretionary{}{}{}\begingroup
  \urlstyle{rm}\Url}\fi

\bibitem[{\textit{Albarede et~al.}(2013)\textit{Albarede, Ballhaus,
  Blichert-Toft, Lee, Marty, Moynier, and Yin}}]{Albarede:2013hd}
Albarede, F., C.~Ballhaus, J.~Blichert-Toft, C.-T. Lee, B.~Marty, F.~Moynier,
  and Q.-Z. Yin (2013), {Asteroidal impacts and the origin of terrestrial and
  lunar volatiles}, \textit{ICARUS}, \textit{222}(1), 44--52.

\bibitem[{\textit{All{\`e}gre et~al.}(2008)\textit{All{\`e}gre, Manh{\`e}s, and
  G{\"o}pel}}]{Allegre:2008iq}
All{\`e}gre, C.~J., G.~Manh{\`e}s, and C.~G{\"o}pel (2008), {The major
  differentiation of the Earth at ∼4.45 Ga}, \textit{Earth and Planetary
  Science Letters}, \textit{267}(1-2), 386--398.

\bibitem[{\textit{Armytage et~al.}(2012)\textit{Armytage, Georg, Williams, and
  Halliday}}]{Armytage:2012br}
Armytage, R. M.~G., R.~B. Georg, H.~M. Williams, and A.~N. Halliday (2012),
  {Silicon isotopes in lunar rocks: Implications for the Moon's formation and
  the early history of the Earth}, \textit{Geochimica et Cosmochimica Acta},
  \textit{77}, 504--514.

\bibitem[{\textit{Asphaug et~al.}(2006)\textit{Asphaug, Agnor, and
  Williams}}]{Asphaug:2006gp}
Asphaug, E., C.~B. Agnor, and Q.~Williams (2006), {Hit-and-run planetary
  collisions}, \textit{Nature}, \textit{439}(7), 155--160.

\bibitem[{\textit{Becker et~al.}(2006)\textit{Becker, Horan, Walker, Gao,
  Lorand, and Rudnick}}]{Becker:2006bi}
Becker, H., M.~F. Horan, R.~J. Walker, S.~Gao, J.~P. Lorand, and R.~L. Rudnick
  (2006), {Highly siderophile element composition of the Earth's primitive
  upper mantle: Constraints from new data on peridotite massifs and xenoliths},
  \textit{Geochimica et Cosmochimica Acta}, \textit{70}(17), 4528--4550.

\bibitem[{\textit{Bottke et~al.}(2010)\textit{Bottke, Walker, Day,
  Nesvorn{\'y}, and Elkins-Tanton}}]{Bottke:2010hr}
Bottke, W.~F., R.~J. Walker, J.~M.~D. Day, D.~Nesvorn{\'y}, and
  L.~Elkins-Tanton (2010), {Stochastic Late Accretion to Earth, the Moon, and
  Mars}, \textit{Science}, \textit{330}(6), 1527--1530.

\bibitem[{\textit{Brasser et~al.}(2013)\textit{Brasser, Walsh, and
  Nesvorn{\'y}}}]{Brasser:2013hs}
Brasser, R., K.~J. Walsh, and D.~Nesvorn{\'y} (2013), {Constraining the
  primordial orbits of the terrestrial planets}, \textit{Monthly Notices of the
  Royal Astronomical Society}, \textit{433}(4), 3417--3427.

\bibitem[{\textit{Canup}(2008)}]{Canup:2008ff}
Canup, R.~M. (2008), {Lunar-forming collisions with pre-impact rotation},
  \textit{ICARUS}, \textit{196}(2), 518--538.

\bibitem[{\textit{Canup}(2012)}]{Canup:2012cd}
Canup, R.~M. (2012), {Forming a Moon with an Earth-like Composition via a Giant
  Impact}, \textit{Science}, \textit{338}(6), 1052.

\bibitem[{\textit{Chambers}(1999)}]{Chambers:1999jd}
Chambers, J.~E. (1999), {A hybrid symplectic integrator that permits close
  encounters between massive bodies}, \textit{Monthly Notices of the Royal
  Astronomical Society}, \textit{304}(4), 793--799.

\bibitem[{\textit{Chambers}(2001)}]{Chambers:2001kt}
Chambers, J.~E. (2001), {Making More Terrestrial Planets}, \textit{ICARUS},
  \textit{152}(2), 205--224.

\bibitem[{\textit{Chambers}(2013)}]{Chambers:2013cp}
Chambers, J.~E. (2013), {Late-stage planetary accretion including hit-and-run
  collisions and fragmentation}, \textit{ICARUS}, \textit{224}(1), 43--56.

\bibitem[{\textit{Chambers and Cassen}(2002)}]{Chambers:2002hj}
Chambers, J.~E., and P.~M. Cassen (2002), {The effects of nebula surface
  density profile and giant-planet eccentricities on planetary accretion in the
  inner solar system}, \textit{Meteoritics {\&} Planetary Science},
  \textit{37}, 1523--1540.

\bibitem[{\textit{Chou}(1978)}]{Chou:1978uu}
Chou, C.~L. (1978), {Fractionation of Siderophile Elements in the Earth's Upper
  Mantle}, \textit{Abstracts of the Lunar and Planetary Science Conference},
  \textit{9}, 219--230.

\bibitem[{\textit{Chyba}(1991)}]{Chyba:1991cq}
Chyba, C.~F. (1991), {Terrestrial mantle siderophiles and the lunar impact
  record}, \textit{ICARUS}, \textit{92}(2), 217--233.

\bibitem[{\textit{Clayton and Mayeda}(1999)}]{Clayton:1999hm}
Clayton, R.~N., and T.~K. Mayeda (1999), {Oxygen isotope studies of
  carbonaceous chondrites}, \textit{Geochimica et Cosmochimica Acta},
  \textit{63}(1), 2089--2104.

\bibitem[{\textit{{\'C}uk and Stewart}(2012)}]{Cuk:2012hj}
{\'C}uk, M., and S.~T. Stewart (2012), {Making the Moon from a Fast-Spinning
  Earth: A Giant Impact Followed by Resonant Despinning}, \textit{Science},
  \textit{338}(6), 1047--1052.

\bibitem[{\textit{Dahl and Stevenson}(2010)}]{Dahl:2010ik}
Dahl, T.~W., and D.~J. Stevenson (2010), {Turbulent mixing of metal and
  silicate during planet accretion --- And interpretation of the Hf-W
  chronometer}, \textit{Earth and Planetary Science Letters}, \textit{295}(1),
  177--186.

\bibitem[{\textit{Dauphas and Marty}(2002)}]{Dauphas:2002hu}
Dauphas, N., and B.~Marty (2002), {Inference on the nature and the mass of
  Earth's late veneer from noble metals and gases}, \textit{Journal of
  Geophysical Research (Planets)}, \textit{107}(E), 5129.

\bibitem[{\textit{Dauphas et~al.}(2004)\textit{Dauphas, Davis, Marty, and
  Reisberg}}]{Dauphas:2004hx}
Dauphas, N., A.~M. Davis, B.~Marty, and L.~Reisberg (2004), {The cosmic
  molybdenum-ruthenium isotope correlation}, \textit{Earth and Planetary
  Science Letters}, \textit{226}(3), 465--475.

\bibitem[{\textit{Day and Walker}(2011)}]{Day:2011uq}
Day, J. M.~D., and R.~J. Walker (2011), {The Highly Siderophile Element
  Composition of the Lunar Mantle}, \textit{Abstracts of the Lunar and
  Planetary Science Conference}, \textit{42}, 1288.

\bibitem[{\textit{Day et~al.}(2007)\textit{Day, Pearson, and
  Taylor}}]{Day:2007bp}
Day, J. M.~D., D.~G. Pearson, and L.~A. Taylor (2007), {Highly Siderophile
  Element Constraints on Accretion and Differentiation of the Earth-Moon
  System}, \textit{Science}, \textit{315}(5), 217.

\bibitem[{\textit{Deguen et~al.}(2011)\textit{Deguen, Olson, and
  Cardin}}]{Deguen:2011ig}
Deguen, R., P.~Olson, and P.~Cardin (2011), {Experiments on turbulent
  metal-silicate mixing in a magma ocean}, \textit{Earth and Planetary Science
  Letters}, \textit{310}(3-4), 303--313.

\bibitem[{\textit{Duncan et~al.}(1998)\textit{Duncan, Levison, and
  Lee}}]{Duncan:1998gn}
Duncan, M.~J., H.~F. Levison, and M.~H. Lee (1998), {A Multiple Time Step
  Symplectic Algorithm for Integrating Close Encounters}, \textit{The
  Astronomical Journal}, \textit{116}(4), 2067--2077.

\bibitem[{\textit{Fischer-G{\"o}dde et~al.}(2011)\textit{Fischer-G{\"o}dde,
  Becker, and Wombacher}}]{FischerGodde:2011jd}
Fischer-G{\"o}dde, M., H.~Becker, and F.~Wombacher (2011), {Rhodium, gold and
  other highly siderophile elements in orogenic peridotites and peridotite
  xenoliths}, \textit{Chemical Geology}, \textit{280}(3), 365--383.

\bibitem[{\textit{Fitoussi and Bourdon}(2012)}]{Fitoussi:2012kp}
Fitoussi, C., and B.~Bourdon (2012), {Silicon Isotope Evidence Against an
  Enstatite Chondrite Earth}, \textit{Science}, \textit{335}(6), 1477.

\bibitem[{\textit{Frost and McCammon}(2008)}]{Frost:2008fb}
Frost, D.~J., and C.~A. McCammon (2008), {The Redox State of Earth's Mantle},
  \textit{Annual Review of Earth and Planetary Sciences}, \textit{36},
  389--420.

\bibitem[{\textit{Frost et~al.}(2004)\textit{Frost, Liebske, Langenhorst,
  McCammon, Tr{\o}nnes, and Rubie}}]{Frost:2004ck}
Frost, D.~J., C.~Liebske, F.~Langenhorst, C.~A. McCammon, R.~G. Tr{\o}nnes, and
  D.~C. Rubie (2004), {Experimental evidence for the existence of iron-rich
  metal in the Earth's lower mantle}, \textit{Nature}, \textit{428}(6),
  409--412.

\bibitem[{\textit{Haisch et~al.}(2001)\textit{Haisch, Lada, and
  Lada}}]{Haisch:2001bx}
Haisch, K. E.~J., E.~A. Lada, and C.~J. Lada (2001), {Disk Frequencies and
  Lifetimes in Young Clusters}, \textit{The Astrophysical Journal},
  \textit{553}(2), L153--L156.

\bibitem[{\textit{Halliday}(2008)}]{Halliday:2008bo}
Halliday, A.~N. (2008), {A young Moon-forming giant impact at 70-110 million
  years accompanied by late-stage mixing, core formation and degassing of the
  Earth}, \textit{Philosophical Transactions of the Royal Society A:
  Mathematical, Physical and Engineering Sciences}, \textit{366}(1883),
  4163--4181.

\bibitem[{\textit{Hansen}(2009)}]{Hansen:2009ke}
Hansen, B. M.~S. (2009), {Formation of the Terrestrial Planets from a Narrow
  Annulus}, \textit{The Astrophysical Journal}, \textit{703}(1), 1131--1140.

\bibitem[{\textit{Jacobsen}(2005)}]{Jacobsen:2005bj}
Jacobsen, S.~B. (2005), {The Hf-W Isotopic System and the Origin of the Earth
  and Moon}, \textit{Annual Review of Earth and Planetary Sciences},
  \textit{33}, 531--570.

\bibitem[{\textit{Kleine et~al.}(2009)\textit{Kleine, Touboul, Bourdon, Nimmo,
  Mezger, Palme, Jacobsen, Yin, and Halliday}}]{Kleine:2009tp}
Kleine, T., M.~Touboul, B.~Bourdon, F.~Nimmo, K.~Mezger, H.~Palme, S.~B.
  Jacobsen, Q.-Z. Yin, and A.~N. Halliday (2009), {Hf-W chronology of the
  accretion and early evolution of asteroids and terrestrial planets},
  \textit{Geochimica et Cosmochimica Acta}, \textit{73}, 5150--5188.

\bibitem[{\textit{Kokubo and Genda}(2010)}]{Kokubo:2010im}
Kokubo, E., and H.~Genda (2010), {Formation of Terrestrial Planets from
  Protoplanets Under a Realistic Accretion Condition}, \textit{The
  Astrophysical Journal Letters}, \textit{714}(1), L21--L25.

\bibitem[{\textit{Kokubo and Ida}(1998)}]{Kokubo:1998ka}
Kokubo, E., and S.~Ida (1998), {Oligarchic Growth of Protoplanets},
  \textit{ICARUS}, \textit{131}(1), 171--178.

\bibitem[{\textit{K{\"o}nig et~al.}(2011)\textit{K{\"o}nig, M{\"u}nker, Hohl,
  Paulick, Barth, Lagos, Pf{\"a}nder, and B{\"u}chl}}]{Konig:2011we}
K{\"o}nig, S., C.~M{\"u}nker, S.~Hohl, H.~Paulick, A.~R. Barth, M.~Lagos,
  J.~Pf{\"a}nder, and A.~B{\"u}chl (2011), {The Earth's tungsten budget during
  mantle melting and crust formation}, \textit{Geochimica et Cosmochimica
  Acta}, \textit{75}, 2119--2136.

\bibitem[{\textit{Laskar}(1997)}]{Laskar:1997vw}
Laskar, J. (1997), {Large scale chaos and the spacing of the inner planets.},
  \textit{Astronomy and Astrophysics}, \textit{317}, L75--L78.

\bibitem[{\textit{Leinhardt and Stewart}(2012)}]{Leinhardt:2012kd}
Leinhardt, Z.~M., and S.~T. Stewart (2012), {Collisions between
  Gravity-dominated Bodies. I. Outcome Regimes and Scaling Laws}, \textit{The
  Astrophysical Journal}, \textit{745}(1), 79.

\bibitem[{\textit{Levison et~al.}(2011)\textit{Levison, Morbidelli, Tsiganis,
  Nesvorn{\'y}, and Gomes}}]{Levison:2011gt}
Levison, H.~F., A.~Morbidelli, K.~Tsiganis, D.~Nesvorn{\'y}, and R.~S. Gomes
  (2011), {Late Orbital Instabilities in the Outer Planets Induced by
  Interaction with a Self-gravitating Planetesimal Disk}, \textit{The
  Astronomical Journal}, \textit{142}(5), 152.

\bibitem[{\textit{Leya et~al.}(2008)\textit{Leya, Sch{\"o}nb{\"a}chler,
  Wiechert, Kr{\"a}henb{\"u}hl, and Halliday}}]{Leya:2008ik}
Leya, I., M.~Sch{\"o}nb{\"a}chler, U.~Wiechert, U.~Kr{\"a}henb{\"u}hl, and
  A.~N. Halliday (2008), {Titanium isotopes and the radial heterogeneity of the
  solar system}, \textit{Earth and Planetary Science Letters},
  \textit{266}(3-4), 233--244.

\bibitem[{\textit{Lugmair and Shukolyukov}(1998)}]{Lugmair:1998iq}
Lugmair, G.~W., and A.~Shukolyukov (1998), {Early solar system timescales
  according to 53Mn-53Cr systematics}, \textit{Geochimica et Cosmochimica
  Acta}, \textit{62}(16), 2863--2886.

\bibitem[{\textit{Maier et~al.}(2009)\textit{Maier, Barnes, Campbell,
  Fiorentini, Peltonen, Barnes, and Smithies}}]{Maier:2009kq}
Maier, W.~D., S.~J. Barnes, I.~H. Campbell, M.~L. Fiorentini, P.~Peltonen,
  S.-J. Barnes, and R.~H. Smithies (2009), {Progressive mixing of meteoritic
  veneer into the early Earth's deep mantle}, \textit{Nature}, \textit{460}(7),
  620--623.

\bibitem[{\textit{Mann et~al.}(2012)\textit{Mann, Frost, Rubie, Becker, and
  Aud{\'e}tat}}]{Mann:2012fp}
Mann, U., D.~J. Frost, D.~C. Rubie, H.~Becker, and A.~Aud{\'e}tat (2012),
  {Partitioning of Ru, Rh, Pd, Re, Ir and Pt between liquid metal and silicate
  at high pressures and high temperatures - Implications for the origin of
  highly siderophile element concentrations in the Earth's mantle},
  \textit{Geochimica et Cosmochimica Acta}, \textit{84}, 593--613.

\bibitem[{\textit{Meisel et~al.}(1996)\textit{Meisel, Walker, and
  Morgan}}]{Meisel:1996dg}
Meisel, T., R.~J. Walker, and J.~W. Morgan (1996), {The osmium isotopic
  composition of the Earth's primitive upper mantle}, \textit{Nature},
  \textit{383}(6600), 517--520.

\bibitem[{\textit{Morbidelli et~al.}(2007)\textit{Morbidelli, Tsiganis, Crida,
  Levison, and Gomes}}]{Morbidelli:2007hy}
Morbidelli, A., K.~Tsiganis, A.~Crida, H.~F. Levison, and R.~S. Gomes (2007),
  {Dynamics of the Giant Planets of the Solar System in the Gaseous
  Protoplanetary Disk and Their Relationship to the Current Orbital
  Architecture}, \textit{The Astronomical Journal}, \textit{134}(5),
  1790--1798.

\bibitem[{\textit{Morbidelli et~al.}(2012{\natexlab{a}})\textit{Morbidelli,
  Marchi, Bottke, and Kring}}]{Morbidelli:2012ko}
Morbidelli, A., S.~Marchi, W.~F. Bottke, and D.~A. Kring (2012{\natexlab{a}}),
  {A sawtooth-like timeline for the first billion years of lunar bombardment},
  \textit{Earth and Planetary Science Letters}, \textit{355}, 144--151.

\bibitem[{\textit{Morbidelli et~al.}(2012{\natexlab{b}})\textit{Morbidelli,
  Lunine, O'Brien, Raymond, and Walsh}}]{Morbidelli:2012iz}
Morbidelli, A., J.~I. Lunine, D.~P. O'Brien, S.~N. Raymond, and K.~J. Walsh
  (2012{\natexlab{b}}), {Building Terrestrial Planets}, \textit{Annual Review
  of Earth and Planetary Sciences}, \textit{40}, 251--275.

\bibitem[{\textit{Morishima et~al.}(2010)\textit{Morishima, Stadel, and
  Moore}}]{Morishima:2010cs}
Morishima, R., J.~Stadel, and B.~Moore (2010), {From planetesimals to
  terrestrial planets: N-body simulations including the effects of nebular gas
  and giant planets}, \textit{ICARUS}, \textit{207}(2), 517--535.

\bibitem[{\textit{O'Brien et~al.}(2006)\textit{O'Brien, Morbidelli, and
  Levison}}]{Obrien:2006jx}
O'Brien, D.~P., A.~Morbidelli, and H.~F. Levison (2006), {Terrestrial planet
  formation with strong dynamical friction}, \textit{ICARUS}, \textit{184}(1),
  39--58.

\bibitem[{\textit{O'Brien et~al.}(2014)\textit{O'Brien, Walsh, Morbidelli,
  Raymond, and Mandell}}]{OBrien:2014bk}
O'Brien, D.~P., K.~J. Walsh, A.~Morbidelli, S.~N. Raymond, and A.~M. Mandell
  (2014), {Water Delivery and Giant Impacts in the 'Grand Tack' Scenario},
  \textit{ICARUS}, \textit{223}, 74--84.

\bibitem[{\textit{Olson and Weeraratne}(2008)}]{Olson:2008hu}
Olson, P., and D.~Weeraratne (2008), {Experiments on metal-silicate plumes and
  core formation}, \textit{Philosophical Transactions of the Royal Society A:
  Mathematical, Physical and Engineering Sciences}, \textit{366}, 4253--4271.

\bibitem[{\textit{Petitat et~al.}(2008)\textit{Petitat, Kleine, Touboul,
  Bourdon, and Wieler}}]{Petitat:2008uj}
Petitat, M., T.~Kleine, M.~Touboul, B.~Bourdon, and R.~Wieler (2008), {Hf-W
  Chronometry of Aubrites and the Evolution of Planetary Bodies},
  \textit{Abstracts of the Lunar and Planetary Science Conference},
  \textit{39}, 2164.

\bibitem[{\textit{Raymond et~al.}(2006)\textit{Raymond, Quinn, and
  Lunine}}]{Raymond:2006kn}
Raymond, S.~N., T.~R. Quinn, and J.~I. Lunine (2006), {High-resolution
  simulations of the final assembly of Earth-like planets I. Terrestrial
  accretion and dynamics}, \textit{ICARUS}, \textit{183}(2), 265--282.

\bibitem[{\textit{Raymond et~al.}(2009)\textit{Raymond, O'Brien, Morbidelli,
  and Kaib}}]{Raymond:2009is}
Raymond, S.~N., D.~P. O'Brien, A.~Morbidelli, and N.~A. Kaib (2009), {Building
  the terrestrial planets: Constrained accretion in the inner Solar System},
  \textit{ICARUS}, \textit{203}(2), 644--662.

\bibitem[{\textit{Raymond et~al.}(2013)\textit{Raymond, Schlichting, Hersant,
  and Selsis}}]{Raymond:2013jg}
Raymond, S.~N., H.~E. Schlichting, F.~Hersant, and F.~Selsis (2013), {Dynamical
  and collisional constraints on a stochastic late veneer on the terrestrial
  planets}, \textit{ICARUS}, \textit{226}(1), 671--681.

\bibitem[{\textit{Reufer et~al.}(2012)\textit{Reufer, Meier, Benz, and
  Wieler}}]{Reufer:2012dz}
Reufer, A., M.~M.~M. Meier, W.~Benz, and R.~Wieler (2012), {A hit-and-run giant
  impact scenario}, \textit{ICARUS}, \textit{221}(1), 296--299.

\bibitem[{\textit{Rubie et~al.}(2003)\textit{Rubie, Melosh, Reid, Liebske, and
  Righter}}]{Rubie:2003hq}
Rubie, D.~C., H.~J. Melosh, J.~E. Reid, C.~Liebske, and K.~Righter (2003),
  {Mechanisms of metal-silicate equilibration in the terrestrial magma ocean},
  \textit{Earth and Planetary Science Letters}, \textit{205}(3), 239--255.

\bibitem[{\textit{Rubie et~al.}(2004)\textit{Rubie, Gessmann, and
  Frost}}]{Rubie:2004fk}
Rubie, D.~C., C.~K. Gessmann, and D.~J. Frost (2004), {Partitioning of oxygen
  during core formation on the Earth and Mars}, \textit{Nature},
  \textit{429}(6), 58--61.

\bibitem[{\textit{Rubie et~al.}(2011)\textit{Rubie, Frost, Mann, Asahara,
  Nimmo, Tsuno, Kegler, Holzheid, and Palme}}]{Rubie:2011cr}
Rubie, D.~C., D.~J. Frost, U.~Mann, Y.~Asahara, F.~Nimmo, K.~Tsuno, P.~Kegler,
  A.~Holzheid, and H.~Palme (2011), {Heterogeneous accretion, composition and
  core-mantle differentiation of the Earth}, \textit{Earth and Planetary
  Science Letters}, \textit{301}(1), 31--42.

\bibitem[{\textit{Rudge et~al.}(2010)\textit{Rudge, Kleine, and
  Bourdon}}]{Rudge:2010fv}
Rudge, J.~F., T.~Kleine, and B.~Bourdon (2010), {Broad bounds on Earth's
  accretion and core formation constrained by geochemical models},
  \textit{Nature Geoscience}, \textit{3}(6), 439--443.

\bibitem[{\textit{Samuel}(2012)}]{Samuel:2012gw}
Samuel, H. (2012), {A re-evaluation of metal diapir breakup and equilibration
  in terrestrial magma oceans}, \textit{Earth and Planetary Science Letters},
  \textit{313}, 105--114.

\bibitem[{\textit{Sharp et~al.}(2013)\textit{Sharp, McCubbin, and
  Shearer}}]{Sharp:2013cp}
Sharp, Z.~D., F.~M. McCubbin, and C.~K. Shearer (2013), {A hydrogen-based
  oxidation mechanism relevant to planetary formation}, \textit{Earth and
  Planetary Science Letters}, \textit{380}, 88--97.

\bibitem[{\textit{Spicuzza et~al.}(2007)\textit{Spicuzza, Day, Taylor, and
  Valley}}]{Spicuzza:2007cb}
Spicuzza, M.~J., J.~M.~D. Day, L.~A. Taylor, and J.~W. Valley (2007), {Oxygen
  isotope constraints on the origin and differentiation of the Moon},
  \textit{Earth and Planetary Science Letters}, \textit{253}(1), 254--265.

\bibitem[{\textit{Strom et~al.}(2005)\textit{Strom, Malhotra, Ito, Yoshida, and
  Kring}}]{Strom:2005ur}
Strom, R.~G., R.~Malhotra, T.~Ito, F.~Yoshida, and D.~A. Kring (2005), {The
  Origin of Planetary Impactors in the Inner Solar System}, \textit{Science},
  \textit{309}(5), 1847--1850.

\bibitem[{\textit{Taylor et~al.}(2009)\textit{Taylor, McKeegan, and
  Harrison}}]{Taylor:2009jr}
Taylor, D.~J., K.~D. McKeegan, and T.~M. Harrison (2009), {Lu-Hf zircon
  evidence for rapid lunar differentiation}, \textit{Earth and Planetary
  Science Letters}, \textit{279}(3), 157--164.

\bibitem[{\textit{Touboul et~al.}(2007)\textit{Touboul, Kleine, Bourdon, Palme,
  and Wieler}}]{Touboul:2007is}
Touboul, M., T.~Kleine, B.~Bourdon, H.~Palme, and R.~Wieler (2007), {Late
  formation and prolonged differentiation of the Moon inferred from W isotopes
  in lunar metals}, \textit{Nature}, \textit{450}(7), 1206--1209.

\bibitem[{\textit{Touboul et~al.}(2012)\textit{Touboul, Puchtel, and
  Walker}}]{Touboul:2012gq}
Touboul, M., I.~S. Puchtel, and R.~J. Walker (2012), {182W Evidence for
  Long-Term Preservation of Early Mantle Differentiation Products},
  \textit{Science}, \textit{335}(6), 1065.

\bibitem[{\textit{Trinquier et~al.}(2008)\textit{Trinquier, Birck, All{\`e}gre,
  G{\"o}pel, and Ulfbeck}}]{Trinquier:2008bp}
Trinquier, A., J.~L. Birck, C.~J. All{\`e}gre, C.~G{\"o}pel, and D.~Ulfbeck
  (2008), {53Mn--53Cr systematics of the early Solar System revisited},
  \textit{Geochimica et Cosmochimica Acta}, \textit{72}(20), 5146--5163.

\bibitem[{\textit{Wade and Wood}(2005)}]{Wade:2005fi}
Wade, J., and B.~J. Wood (2005), {Core formation and the oxidation state of the
  Earth}, \textit{Earth and Planetary Science Letters}, \textit{236}(1),
  78--95.

\bibitem[{\textit{Walker}(2009)}]{Walker:2009be}
Walker, R.~J. (2009), {Highly siderophile elements in the Earth, Moon and Mars:
  Update and implications for planetary accretion and differentiation},
  \textit{Chemie der Erde - Geochemistry}, \textit{69}, 101--125.

\bibitem[{\textit{Walker et~al.}(2004)\textit{Walker, Horan, Shearer, and
  Papike}}]{Walker:2004hs}
Walker, R.~J., M.~F. Horan, C.~K. Shearer, and J.~J. Papike (2004), {Low
  abundances of highly siderophile elements in the lunar mantle: evidence for
  prolonged late accretion}, \textit{Earth and Planetary Science Letters},
  \textit{224}(3), 399--413.

\bibitem[{\textit{Walsh et~al.}(2011)\textit{Walsh, Morbidelli, Raymond,
  O'Brien, and Mandell}}]{Walsh:2011co}
Walsh, K.~J., A.~Morbidelli, S.~N. Raymond, D.~P. O'Brien, and A.~M. Mandell
  (2011), {A low mass for Mars from Jupiter's early gas-driven migration},
  \textit{Nature}, \textit{475}(7), 206--209.

\bibitem[{\textit{Wang and Becker}(2013)}]{Wang:2013iy}
Wang, Z., and H.~Becker (2013), {Ratios of S, Se and Te in the silicate Earth
  require a volatile-rich late veneer}, \textit{Nature}, \textit{499}(7),
  328--331.

\bibitem[{\textit{Wetherill}(1991)}]{Wetherill:1991wc}
Wetherill, G.~W. (1991), {Why Isn't Mars as Big as Earth?}, \textit{Abstracts
  of the Lunar and Planetary Science Conference}, \textit{22}, 1495.

\bibitem[{\textit{Wiechert et~al.}(2001)\textit{Wiechert, Halliday, Lee,
  Snyder, Taylor, and Rumble}}]{Wiechert:2001il}
Wiechert, U., A.~N. Halliday, D.-C.~C. Lee, G.~A. Snyder, L.~A. Taylor, and
  D.~Rumble (2001), {Oxygen Isotopes and the Moon-Forming Giant Impact},
  \textit{Science}, \textit{294}(5), 345--348.

\bibitem[{\textit{Willbold et~al.}(2011)\textit{Willbold, Elliott, and
  Moorbath}}]{Willbold:2011kw}
Willbold, M., T.~Elliott, and S.~Moorbath (2011), {The tungsten isotopic
  composition of the Earth's mantle before the terminal bombardment},
  \textit{Nature}, \textit{477}(7363), 195--198.

\bibitem[{\textit{Yin et~al.}(2002)\textit{Yin, Jacobsen, Yamashita,
  Blichert-Toft, T{\'e}louk, and Albar{\`e}de}}]{Yin:2002ex}
Yin, Q., S.~B. Jacobsen, K.~Yamashita, J.~Blichert-Toft, P.~T{\'e}louk, and
  F.~Albar{\`e}de (2002), {A short timescale for terrestrial planet formation
  from Hf--W chronometry of meteorites}, \textit{Nature}, \textit{418}(6901),
  949--952.

\bibitem[{\textit{Zhang et~al.}(2012)\textit{Zhang, Dauphas, Davis, Leya, and
  Fedkin}}]{Zhang:2012dn}
Zhang, J., N.~Dauphas, A.~M. Davis, I.~Leya, and A.~Fedkin (2012), {The
  proto-Earth as a significant source of lunar material}, \textit{Nature
  Geoscience}, \textit{5}(4), 251--255.

\end{thebibliography}
\bibliographystyle{agufull08.bst}

\end{document}